\documentclass[journal]{IEEEtran}
\usepackage{mathptmx}
\usepackage{amsmath}
\usepackage{amssymb}
\usepackage[ruled] {algorithm2e}
\usepackage{algpseudocode}
\usepackage{subcaption}
\usepackage[flushleft]{threeparttable}
\usepackage{booktabs}
\usepackage{color}
\usepackage{soul}
\usepackage{enumitem}
\usepackage{multirow}
\usepackage{bigstrut}
\usepackage{graphicx}
\usepackage{bm}
\usepackage{array} 
\usepackage{hyperref}
\usepackage{orcidlink}

\DeclareSymbolFont{newfont}{OML}{cmm}{m}{it}
\DeclareMathAlphabet{\mathcal}{OMS}{cmsy}{m}{n}
\DeclareMathSymbol{\Varrho}{3}{newfont}{37}

\algnewcommand\And{\textbf{and}}

\allowdisplaybreaks

\begin{document}
\title{Device-Free Localization Using Commercial UWB Transceivers}

\author{
Hyun Seok Lee\,\orcidlink{0009-0005-7999-1832},~LG Innotek, Seoul, South Korea%
\thanks{H.~S.~Lee is with the Convenience Control Connectivity S/W Development Team, LG Innotek, Seoul 07796, Korea (e-mail: hyunseok1.lee@lginnotek.com, dlgustjr0443@snu.ac.kr).}%
\thanks{This work was conducted during the author's Master's program at the Dept. of Computer Science and Engineering, Seoul National University, Korea. The author is currently employed at LG Innotek, but the research was completed prior to joining the company.

This initial manuscript reflects results obtained before collaboration with other authors, and an improved version of this work may exist as part of joint research.}%
}
\maketitle

\begin{abstract}
Recently, commercial ultra-wideband (UWB) transceivers have enabled not only measuring device-to-device distance but also tracking the position of a pedestrian who does not carry a UWB device. 
UWB-based device-free localization that does not require dedicated radar equipment is compatible with existing anchor infrastructure and can be reused to reduce hardware deployment costs. 
However, it is difficult to estimate the target's position accurately in real-world scenarios due to the low signal-to-noise ratio (SNR) and the cluttered  environment. 
 In this paper, we propose a deep learning (DL)-assisted particle filter to overcome these challenges. 
First, the channel impulse response (CIR) variance is analyzed to capture the variability induced by the target’s movement.
Then, a DL-based 1-D attention U-Net is used to extract only the reflection components caused by the target and suppress the noise components within the CIR variance profile. 
Finally, the multiple preprocessed CIR variance profiles are used as input to a particle filter to estimate the target's position. 
Experimental results demonstrate that the proposed system is a practical and cost-effective solution for IoT and automotive applications with a root mean square error (RMSE) of about 15~cm and an average processing time of 4~ms. 
Furthermore, comparisons with existing state-of-the-art methods show that the proposed method provides the best performance with reasonable computational costs.
\end{abstract}

\begin{IEEEkeywords}
ultra-wideband (UWB), Device-free localization (DFL), channel impulse response (CIR), Region of interest estimation with deep learning (DL), DL-assisted particle filter.
\end{IEEEkeywords}
\IEEEpeerreviewmaketitle

\begin{figure*}[!t] 

\begin{subfigure}[t]{0.24\textwidth}
\includegraphics[width=\textwidth]{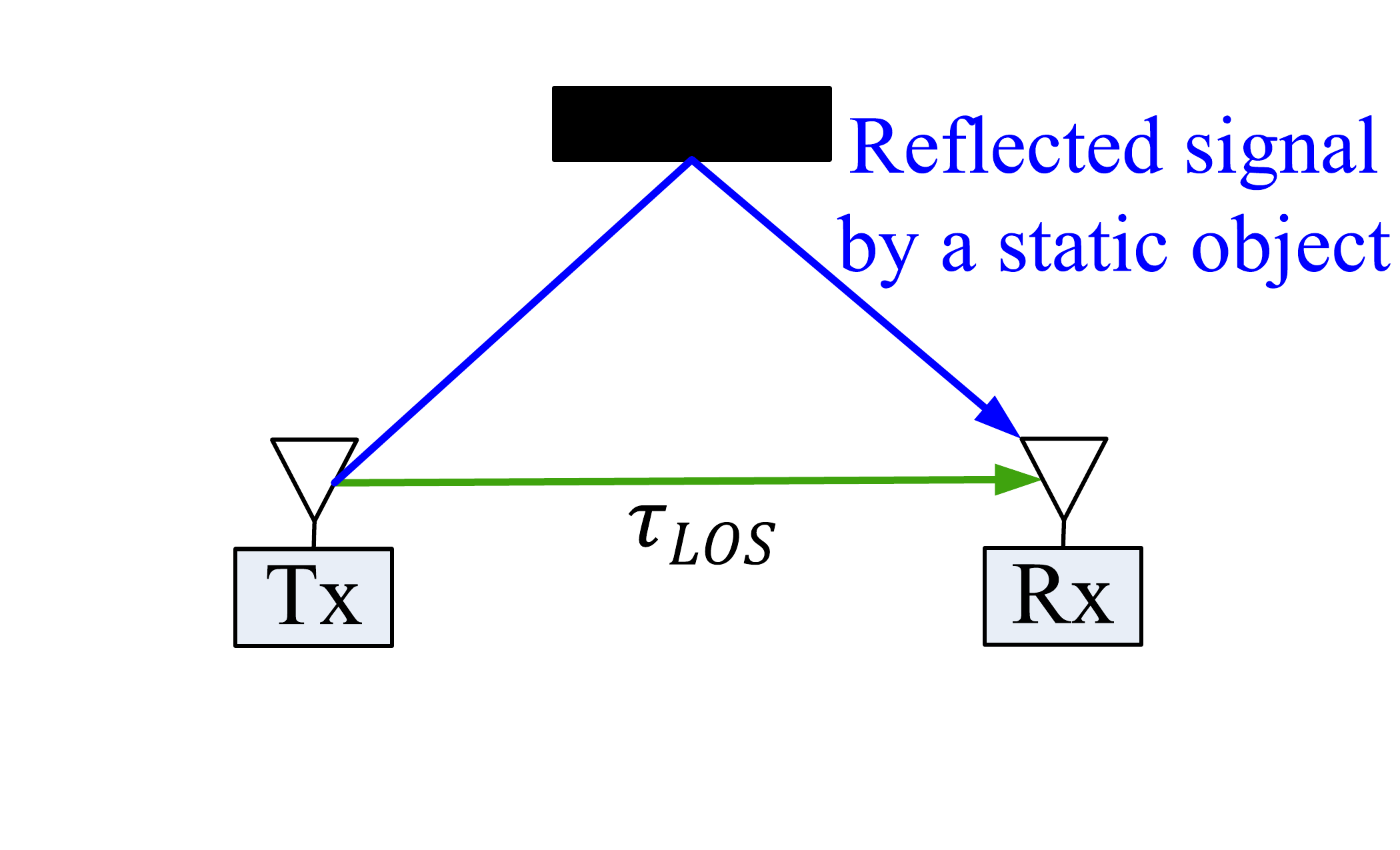}
\vspace{-1.5em} 			
\caption{Background scenario.}
\end{subfigure}
 \hspace{-0.8em}
\begin{subfigure}[t]{0.24\textwidth}
\includegraphics[width=\textwidth]{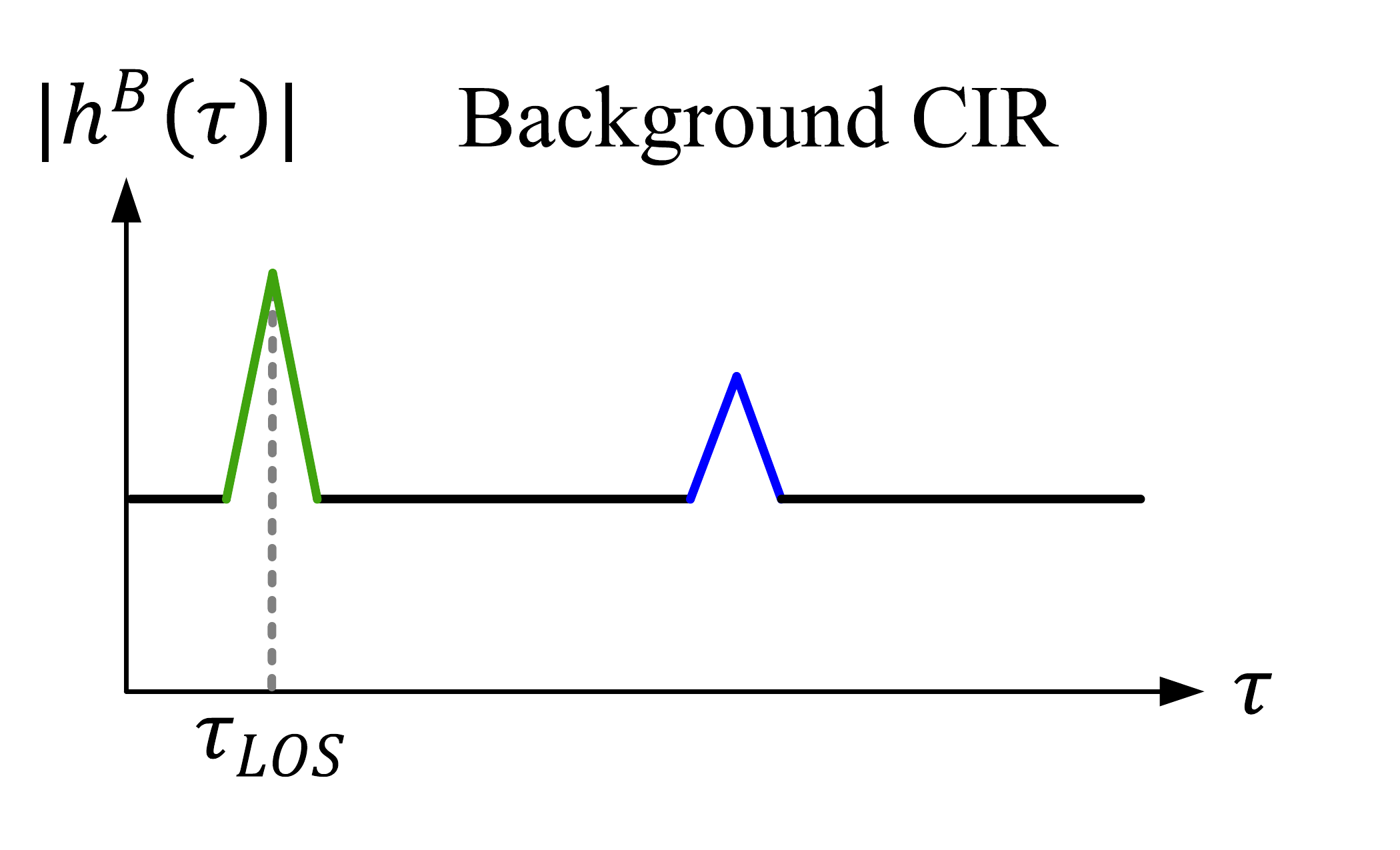}
\vspace{-1.5em} 
\caption{Ideal background CIR.}
\end{subfigure}
\hspace{-1em}
\begin{subfigure}[t]{0.25\textwidth}
\raisebox{0.3em}{    \includegraphics[width=\textwidth]{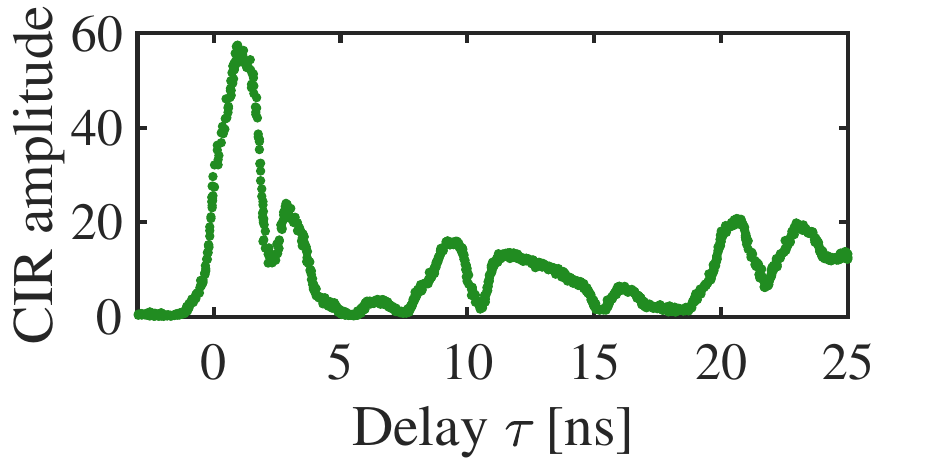}   }
\vspace{-1.5em}  
\caption{Measured background CIR.}
\end{subfigure}
\hspace{-1em}
\begin{subfigure}[t]{0.25\textwidth}
\raisebox{0.3em}{    \includegraphics[width=\textwidth]{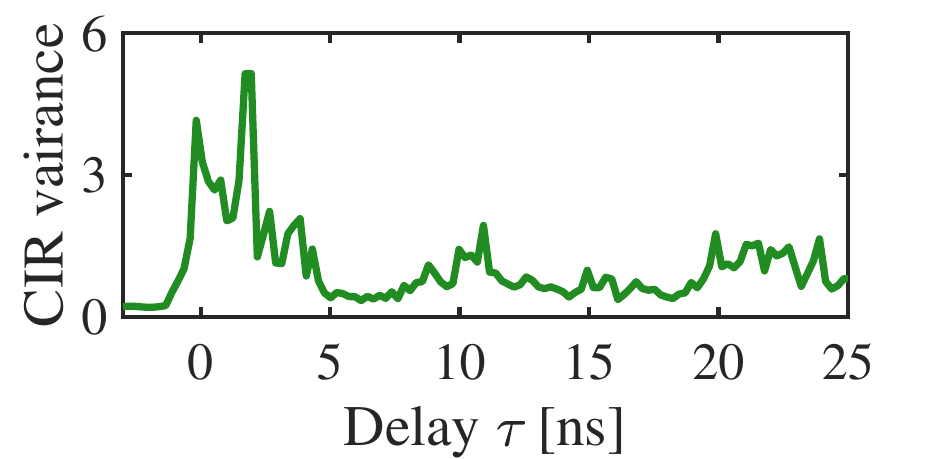}   }
\vspace{-1.5em} 
\caption{Background CIR variance.}
\end{subfigure}

\vspace{-0.5em}
    
\begin{subfigure}[t]{0.24\textwidth}
\includegraphics[width=\textwidth]{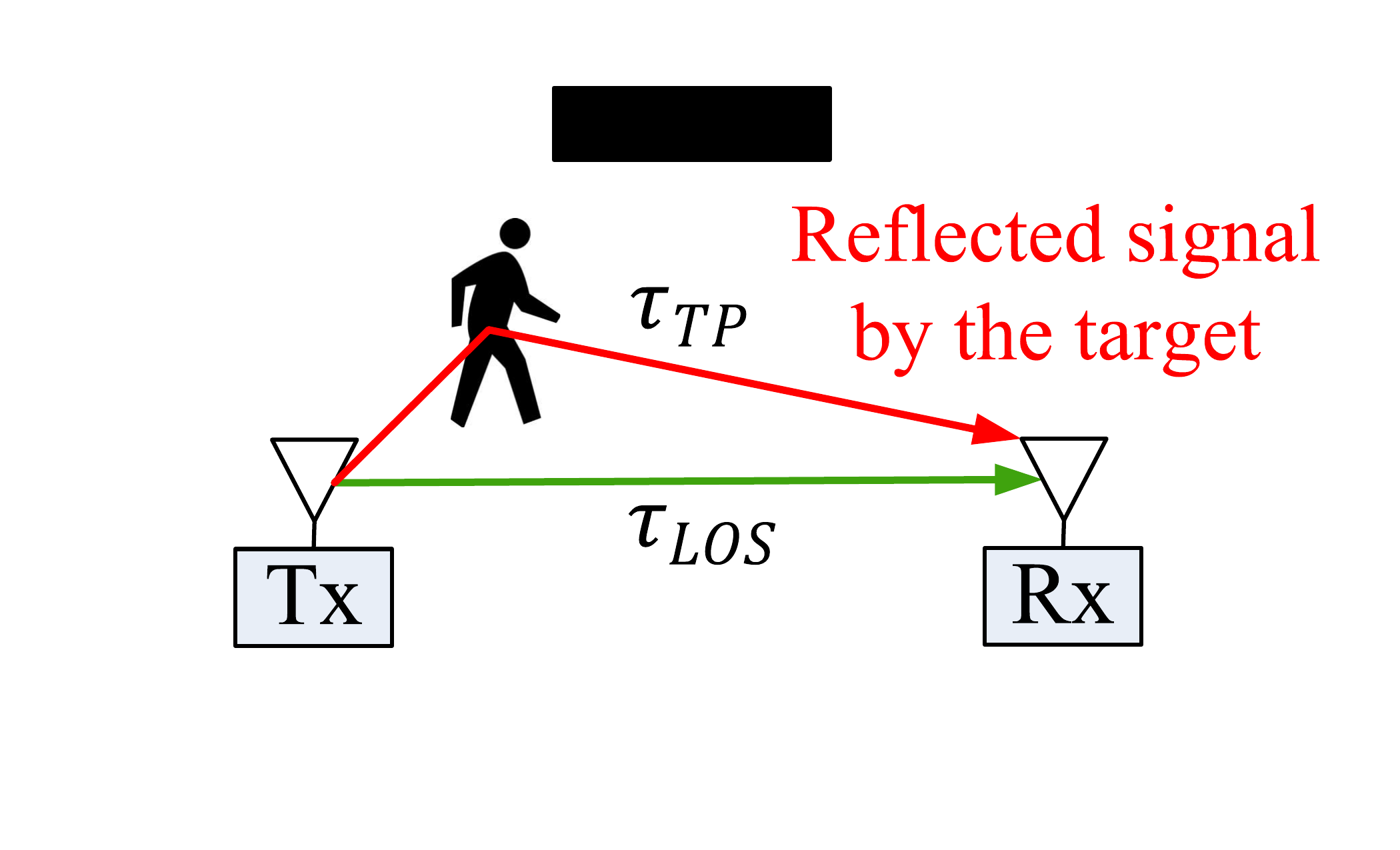}
\vspace{-1.5em} 			
\caption{Dynamic scenario.}
\end{subfigure}
 \hspace{-0.8em}
\begin{subfigure}[t]{0.24\textwidth}
\includegraphics[width=\textwidth]{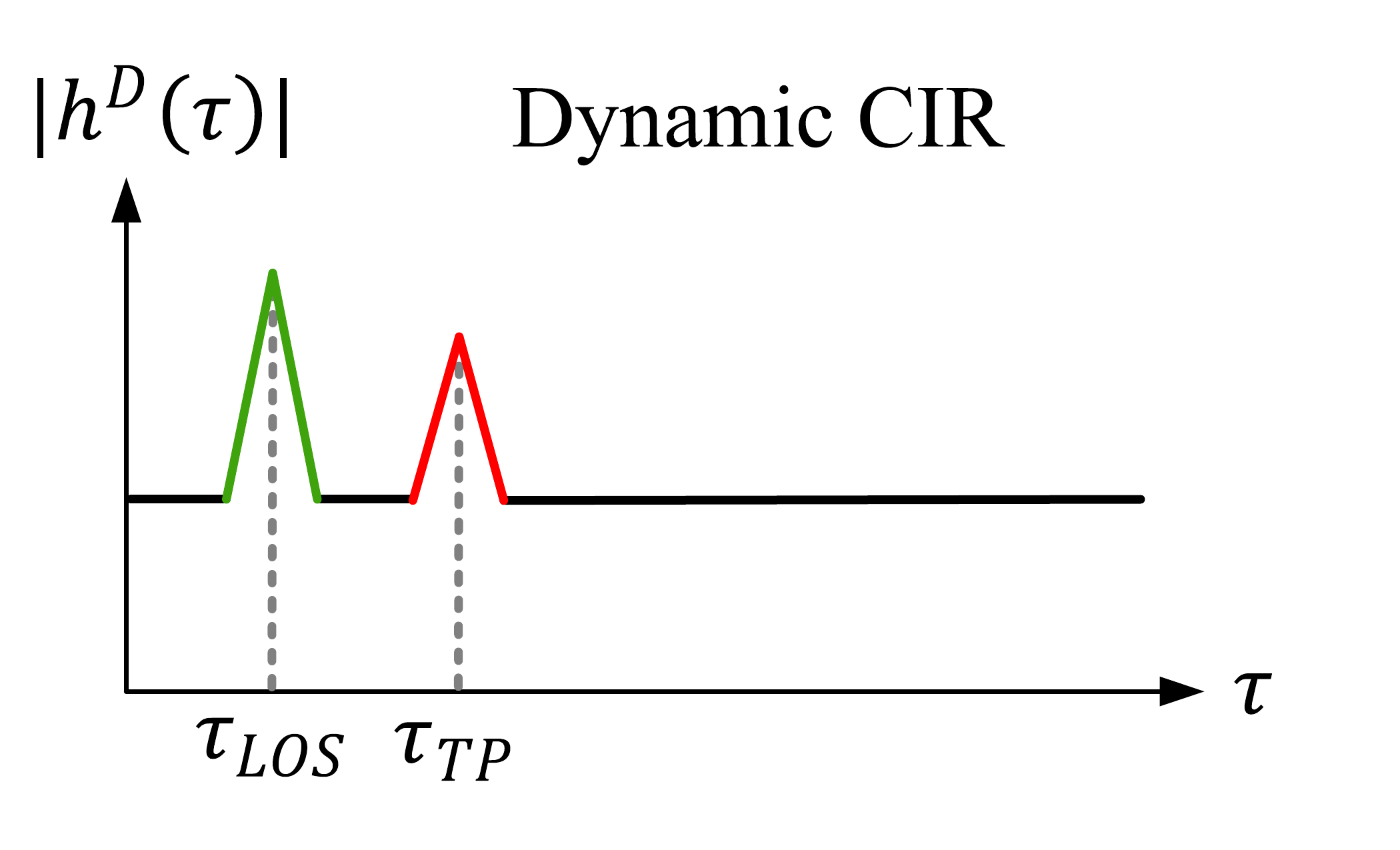}
\vspace{-1.5em} 
\caption{Ideal dynamic CIR.}
\end{subfigure}
\hspace{-1em}
\begin{subfigure}[t]{0.25\textwidth}
\raisebox{0.3em}{    \includegraphics[width=\textwidth]{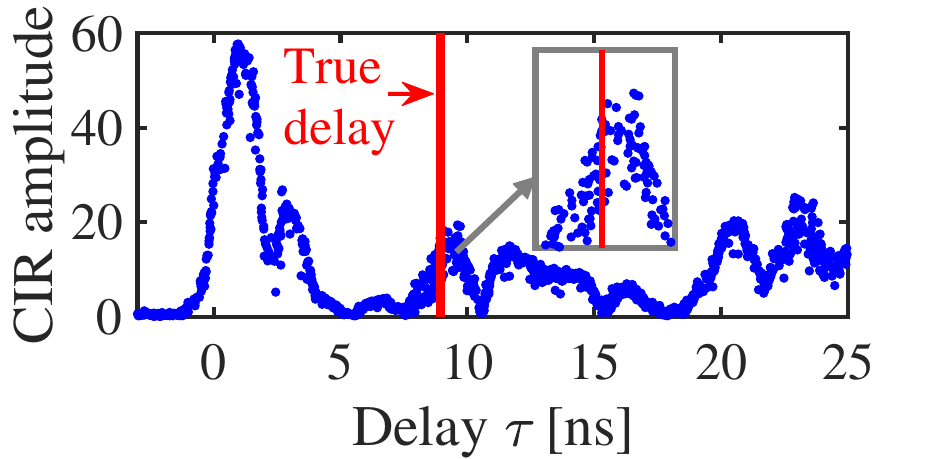}   }
\vspace{-1.5em}  
\caption{Measured dynamic CIR.}
\end{subfigure}
\hspace{-1em}
\begin{subfigure}[t]{0.25\textwidth}
\raisebox{0.3em}{    \includegraphics[width=\textwidth]{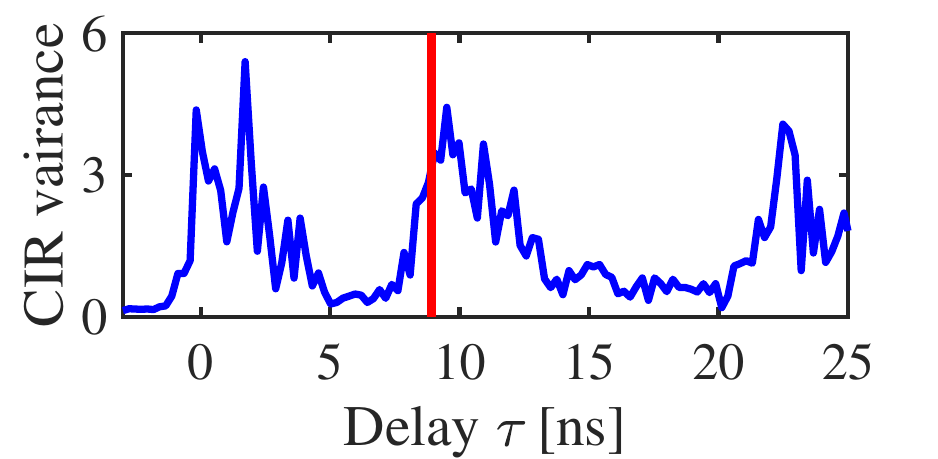}   }
\vspace{-1.5em} 
\caption{Dynamic CIR variance.}
\end{subfigure}
\label{concpet_DFL} 
\vspace{-0.5em}

\begin{subfigure}[t]{0.24\textwidth}
\includegraphics[width=\textwidth]{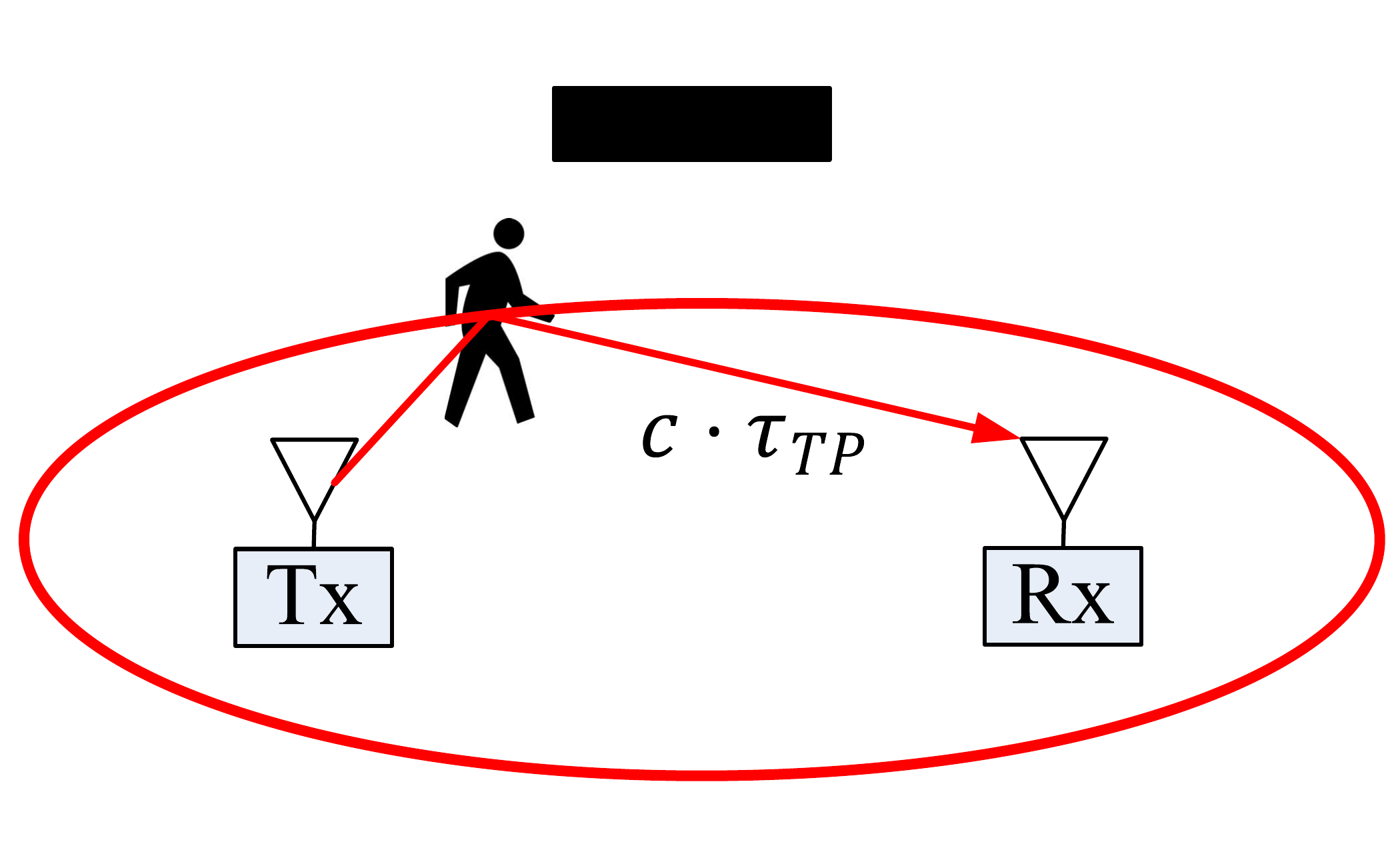}
\vspace{-1.5em} 			
\caption{Generated an ellipse.}
\end{subfigure}
 \hspace{-0.8em}
\begin{subfigure}[t]{0.24\textwidth}
\includegraphics[width=\textwidth]{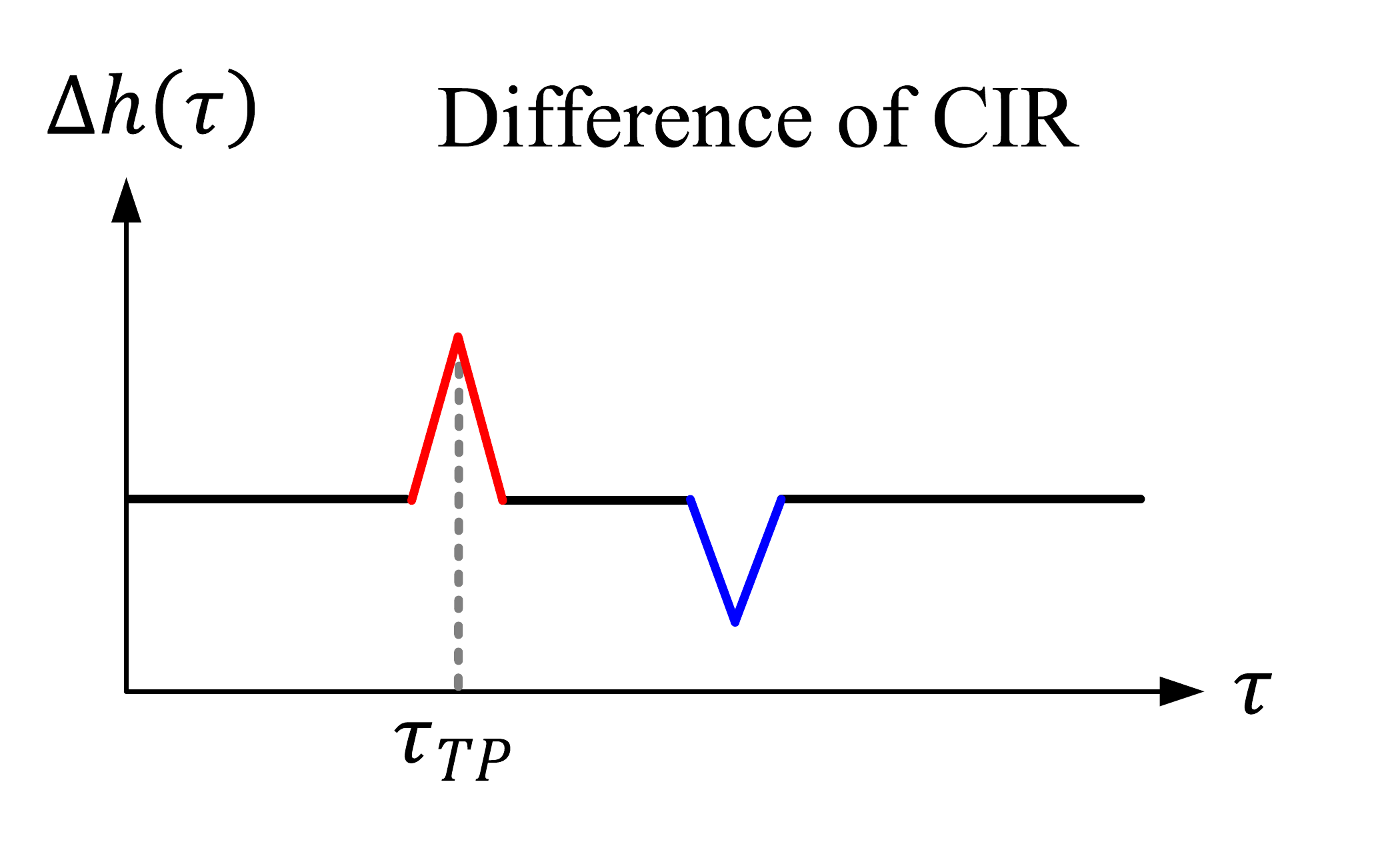}
\vspace{-1.5em} 
\caption{Ideal difference of CIR.}
\end{subfigure}
\hspace{-1em}
\begin{subfigure}[t]{0.25\textwidth}
\raisebox{0.3em}{    \includegraphics[width=\textwidth]{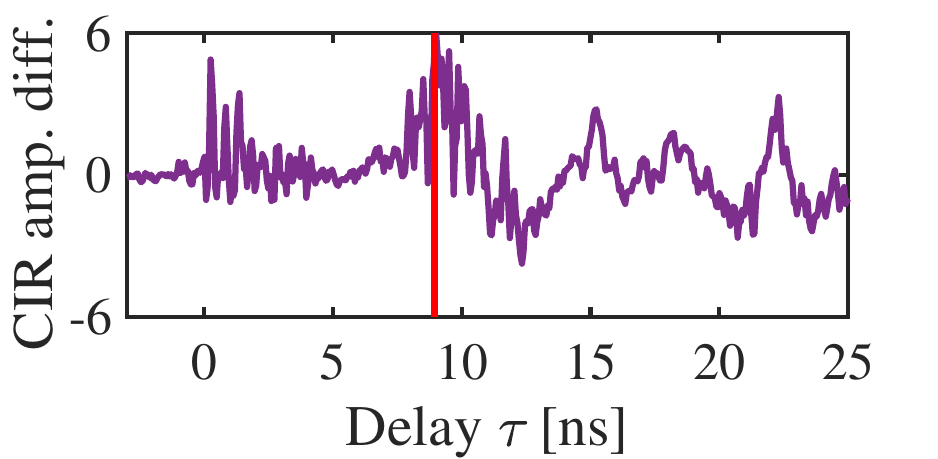}   }
\vspace{-1.5em}  
\caption{Measured difference of CIR.}
\end{subfigure}
\hspace{-1em}
\begin{subfigure}[t]{0.25\textwidth}
\raisebox{0.3em}{    \includegraphics[width=\textwidth]{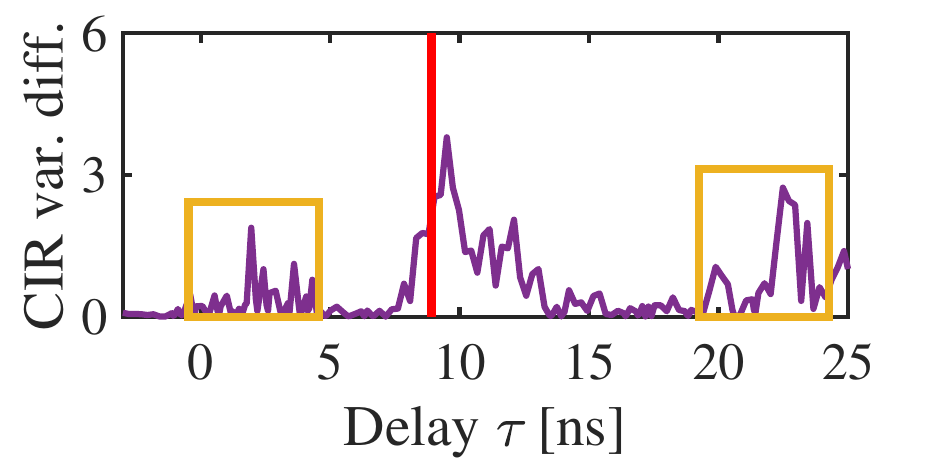}   }
\vspace{-1.5em} 
\caption{Difference of CIR variance.}
\end{subfigure}

 \caption{The concept of the UWB-based DFL.}
\end{figure*}

\section{Introduction}
Radio frequency (RF)-based localization systems are becoming increasingly popular as the advance of the Internet of Things (IoT) and connectivity solutions for automotive. 
According to the Fine Ranging (FiRa) consortium \cite{FiRa:UWB}, accurate positional information enables various applications such as hands-free access control for IoT devices, vehicle digital keys, patient tracking, virtual reality/augmented reality, and finding someone/something nearby services. 

The localization systems based on RF mainly use Bluetooth low energy, radio frequency identification, Wi-Fi, and ultra-wideband (UWB). 
Among these, UWB supports high-resolution ranging information due to its wide frequency bandwidth, which makes it easy to distinguish between line-of-sight (LOS) and multipath components. 

UWB-based localization systems can be categorized into active and passive modes. 
In active mode, a target (e.g., a person or robot) is equipped with an electronic device that communicates with the infrastructure. 
The infrastructure analyzes the received signals to estimate the target’s position.
For signal analysis, received signal strength, time of flight, time difference of arrival, and angle of arrival using phase difference in an antenna array are commonly used \cite{Jeon:2024}. 

On the other hand, passive mode localization systems track the position of a target who does not carry any device. 
Also known as device-free localization (DFL) systems, these systems typically estimate a target's position by extracting the reflected signals caused by the target from the channel impulse response (CIR). 
The CIR captures the wireless channel characteristics including the LOS and multipath components such as reflections, diffractions, and scatterings by the target and environment. 
An overview of UWB-based DFL is provided in \cite{Cheraghinia:2024}. 

UWB-based DFL can be helpful in many use cases such as seamless smart home, elderly care, vehicle theft detection, and rescue scenarios. 
In practice, this is attracting significant industry interest. 
The presence detection capabilities currently under discussion in the IEEE 802.15.4ab standard \cite{IEEE:802.15.4ab-2021} reflect demands from UWB consortiums such as the FiRa consortium \cite{FiRa:UWB} and the Car Connectivity Consortium (CCC) \cite{CCC:UWB}. 
In addition, LG Innotek has recently developed a next-generation UWB digital key solution with radar capabilities for presence detection \cite{LG:Innotek}.

Despite this industry interest, most research on UWB-based DFL has primarily focused on distance estimation, which is the bistatic range between UWB devices and the target \cite{Santoro:2023}, \cite{Herbruggen:2023}.  
This method alone is not sufficient for typical scenarios that require positional data. 

Recently, the multistatic-based DFL utilizing multiple UWB transceivers has been proposed but still faces challenges. 
In \cite{Ledergerber:2020}, a leading-edge detection algorithm was proposed to estimate the propagation delay caused by the target and to achieve a positioning error with a mean square error (RMSE) of 0.33~m. 
However, it is sensitive to environmental noise, making it difficult to maintain consistent accuracy. 
To overcome this limitation of the traditional method, a recent study designed deep learning (DL)-based convolutional neural networks (CNN) that leverage CIR for target path delay estimation via regression and achieve a positioning error with the RMSE of 0.27~m \cite{Li:2022:CNN}. 
This method can reduce the impact of noise and multipath interference but often shows degraded performance in low signal-to-noise ratio (SNR), where detecting target signal components becomes challenging. 

In \cite{Li:2022:MAP}, an alternative approach to improve localization accuracy was proposed by mapping the variance profiles of the CIR from the propagation delay domain to the Cartesian coordinates and estimating the position via approximate maximum a posteriori (MAP). 
It achieves a 90th percentile positional error of 27~cm but is computationally intensive, which limits its use for real-time applications.

To address these challenges, we propose a DL-assisted particle filter for UWB-based DFL. 
The proposed method begins by analyzing the variance profile of the CIR to detect the weak signals reflected by the target. 
Then, a DL-based 1-D attention U-Net is used to predict the region of interest (RoI) where the target-induced signals are likely to be found in the variance profile of the CIR. 
By simplifying the task scope to classification rather than regression, this approach makes it can achieve reliable performance even under low SNR conditions. 
The predicted RoI is used to mask the variance profile of the CIR to extract only the signals relevant to the target. 
Subsequently, the multiple masked variance profiles of the CIR are used as input to a particle filter to estimate the target's position. 
The proposed system achieves an RMSE of about 15~cm and a 90th-percentile error of about 24~cm with an average processing time of 4~ms using Qorvo's DWM1000 commercial UWB transceivers.

This paper is organized as follows: Section II describes the foundational concept of UWB-based DFL. 
Section III details the proposed DL-assisted particle filter. 
Section IV provides experimental validation of the system’s performance, and this paper concludes in Section V.
Section VI discusses the current limitations of the proposed approach and outlines directions for future work.

\section{Preliminaries}
In this section, we describe the concept of estimating the position by extracting the signal reflected by the target from the CIR.
\subsection{Channel Impulse Response}
In wireless communication systems, the CIR characterizes the propagated signals between a transmitter (Tx) and a receiver (Rx). 
It captures the LOS signal, which travels a direct ray between the Tx and the Rx, and reflected rays caused by static and dynamic objects in the environment. 
The time-varying CIR  $h(\tau, t)$ at time $t$ can be expressed as:

\begin{equation}
h(\tau, t) = \sum_j a_j(t) \delta(\tau - \tau_j(t)) + n(\tau, t),
\label{CIR}
\end{equation}

where $a_j(t)$ denotes the channel gain of $j$-th ray, $\delta(\cdot)$ is the Dirac delta function, $\tau_j(t)$ is the propagation delay of $j$-th ray (i.e., the arrival time of the ray), and  $n(\tau, t)$ is the additive measurement noise.
Commercial UWB transceivers use a ternary Ipatov sequence with perfect periodic autocorrelation properties as the preamble code and estimate the CIR based on the received preamble symbols.

As shown in Fig.~1 (a) and (e), the CIR can be categorized into two types of measurement scenarios: the background CIR $h_B(\tau)$ and the dynamic CIR $h_D(\tau, t)$. 
The background CIR $h_B(\tau)$ represents the stationary case including the LOS signal and the reflections by the static object(s). 
The LOS signal (green) dominates the CIR and a static ray (blue) is caused by the surrounding object in the environment, as depicted in Fig.~1(b).

When a target comes into the sensing area shown in Fig.~1(e), additional time-varying reflection component(s) in the CIR are introduced. 
As depicted in Fig.~1(f), a target-induced reflection (red) appears as an additional propagation delay in the CIR, representing the dynamic movement of the target. 
In this paper, we refer to the signal obtained in this scenario as dynamic CIR.  

To extract the target-induced reflections, the background CIR $h_B(\tau)$ is subtracted from the dynamic CIR $h_D(\tau, t)$:
\begin{equation}
\Delta h(\tau, t) = h_D(\tau, t) - h_B(\tau).
\label{CIR difference}
\end{equation}

This subtraction mitigates the background impacts (also known as clutter) and highlights the target-induced signals. 
As shown in Fig.~1(j), the difference of the CIR $\Delta h(\tau, t)$ shows that the propagation delay of the main component is caused by the moving target. 
This delay for the target path can be formulated as $\tau_{TP} = \tau_{Tx} + \tau_{Rx}$. 
It represents the arrival time of the signal that has traveled from the Tx to the target and then to the Rx.  
Then, it can be converted from the propagation delay domain to the spatial domain by multiplying the speed of light $c$, which forms an ellipse with the Tx and the Rx as the focal points shown in Fig.~1(i). 
If multiple transceivers are deployed in the sensing area, each Tx-Rx pair yields an ellipse independently. 
Finally, we can obtain the target's position using the intersection point(s) of the ellipses.

\subsection{Variance-based Analysis for Target Detection}
In practice, detecting target-induced reflections in the CIR is difficult when the target is weakly reflective or the distance between the devices and the target is large. 
Under such conditions, the target-induced signal is often cluttered by noise or background reflections, as shown in Fig.~1(k).

The variance of the CIR over a short segment in the propagation delay domain is a useful analysis for mitigating these issues~\cite{Ledergerber:2020}. 
As presented in Fig.~1(l), the main component in the difference of the CIR variance aligns closely with the true delay caused by the target. 

However, the variance-based CIR analysis still struggles to extract target-induced signals due to residual noise components [see orange boxes in Fig.~1 (l)]. 
First, measurement noise arises from LOS variations caused by hardware imperfections. 
In addition, irrelevant variations occur due to environmental reflections from surrounding static objects.

In this paper, we use a 1-D attention U-Net to predict the RoI in the variance profile of the CIR to extract only target-related reflections. 
After that, the predicted RoI is applied as signal masking to the CIR variance profile to remove residual noise components.

\section{Proposed method}
\subsection{System Overview}

\vspace{2mm}
\begin{figure}[t]\centering
\vspace{2mm}
\includegraphics[width=8.8cm] {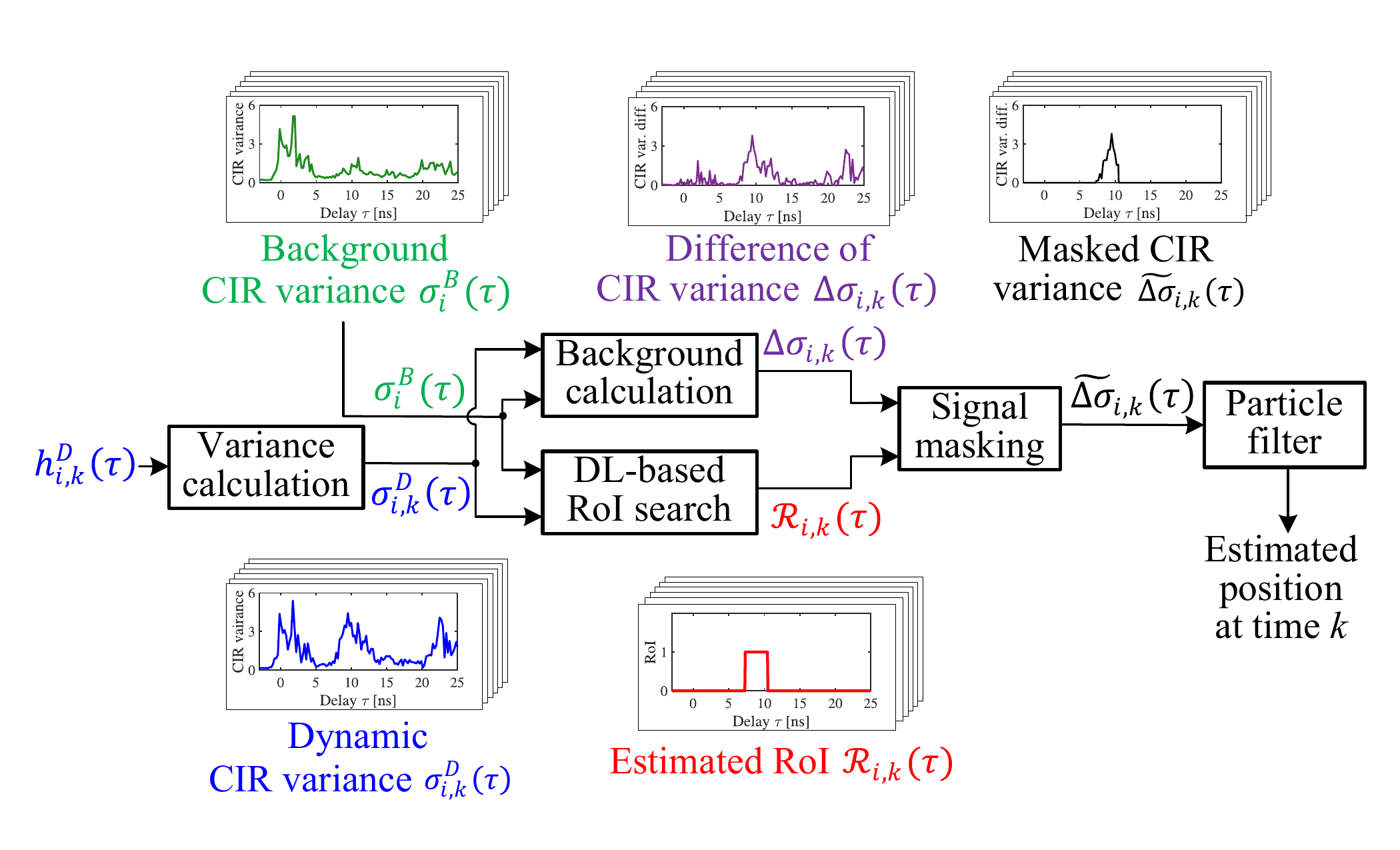}
\vspace{-2mm}
\caption{Proposed system oveview.}\centering
\vspace{-1mm}
\end{figure}

The proposed system estimates the target's position by combining DL-based RoI prediction and a particle filter framework. 
As shown in Fig.~2, the processing starts by calculating the variance of the CIR for both the background and dynamic scenarios to capture the temporal characteristics to analyze the target’s movement. 
Since the background CIR does not change over time, it is calculated only once beforehand and used as a fixed value for each channel link. 
Then, the background CIR is subtracted from the dynamic CIR to extract the signal by the moving target. 
However, the noise components make it hard to pinpoint the target path delay.

To address this, a 1-D attention U-Net is used to identify the RoI likely to contain the boundaries of the target reflections. 
After that, the signal masking is applied to emphasize the difference of the CIR variance within the detected RoI, suppressing irrelevant noise components.
Note that the system processes data from all possible channel links $N_\text{UWB}^\text{Pair}$. 
Finally, the multiple masked CIR variances are used in a particle filter to estimate the target's position from a Bayesian perspective. 

In this paper, the subscript $i$ refers to the channel link pair index of the transceivers, and $k$ is the measurement time index.

\subsection{CIR Variance Calculation}
The variance calculation is used to utilize the time-varying characteristic of the CIR to quantify fluctuations caused by the moving target. 
The background CIR variance $\sigma_i^B(\tau)$ and dynamic CIR variance $\sigma_{i,k}^D(\tau)$ are calculated using consecutive CIR measurements for each scenario.

To be specific about the implementation, the raw CIR delay resolution of 1~ns (30~cm spatially) is insufficient for capturing fine variations of the target. 
Also, the oscillator mismatch between the Tx and the Rx causes a sampling offset. 
To resolve these issues, the raw CIR is aligned to the first path delay reported by the UWB transceiver~\cite{Qorvo:2017}, which has 64 times finer resolution. 
Then, the variance profile is calculated within a short propagation delay bin to capture localized fluctuations in the CIR profile, as shown in Fig.~1(d) and (h). 

The calculated CIR variance profile has a resolution of 0.059~ns and a data size of 500. 
Note that the background CIR variance $\sigma_i^B(\tau)$ is determined by the environment and is time-invariant, it does not require periodic updates and is calculated only once. 
This CIR obtained in the stationary scenario represents noise power in each propagation delay component, which helps to extract target-induced reflections. 
For further details on variance calculation, please refer to~\cite{Li:2022:CNN}.

\subsection{Background Subtraction}
Background subtraction is the process of subtracting the stationary components in the CIR variance to extract the target-induced reflections.
The background CIR variance $\sigma_i^B(\tau)$ is subtracted from the dynamic CIR variance $\sigma_{i,k}^D(\tau)$ to calculate the difference of CIR variance $\Delta \sigma_{i,k}(\tau)$ using (2):

\begin{equation}
\Delta \sigma_{i,k}(\tau) = \sigma_{i,k}^D(\tau) - \sigma_i^B(\tau).
\label{Difference of the CIR variance}
\end{equation}

As presented in Fig.~1(l), this process emphasizes fluctuations induced by the target but introduces some noise components as described in the preliminary. 
The following step discusses how to mitigate these in subsequent processing.

\subsection{Deep Learning-based Region of Interesting Search}

\begin{figure}
\begin{subfigure}{.2\textwidth}
\hspace*{-0.5cm}
\centering
\includegraphics[width=4cm]{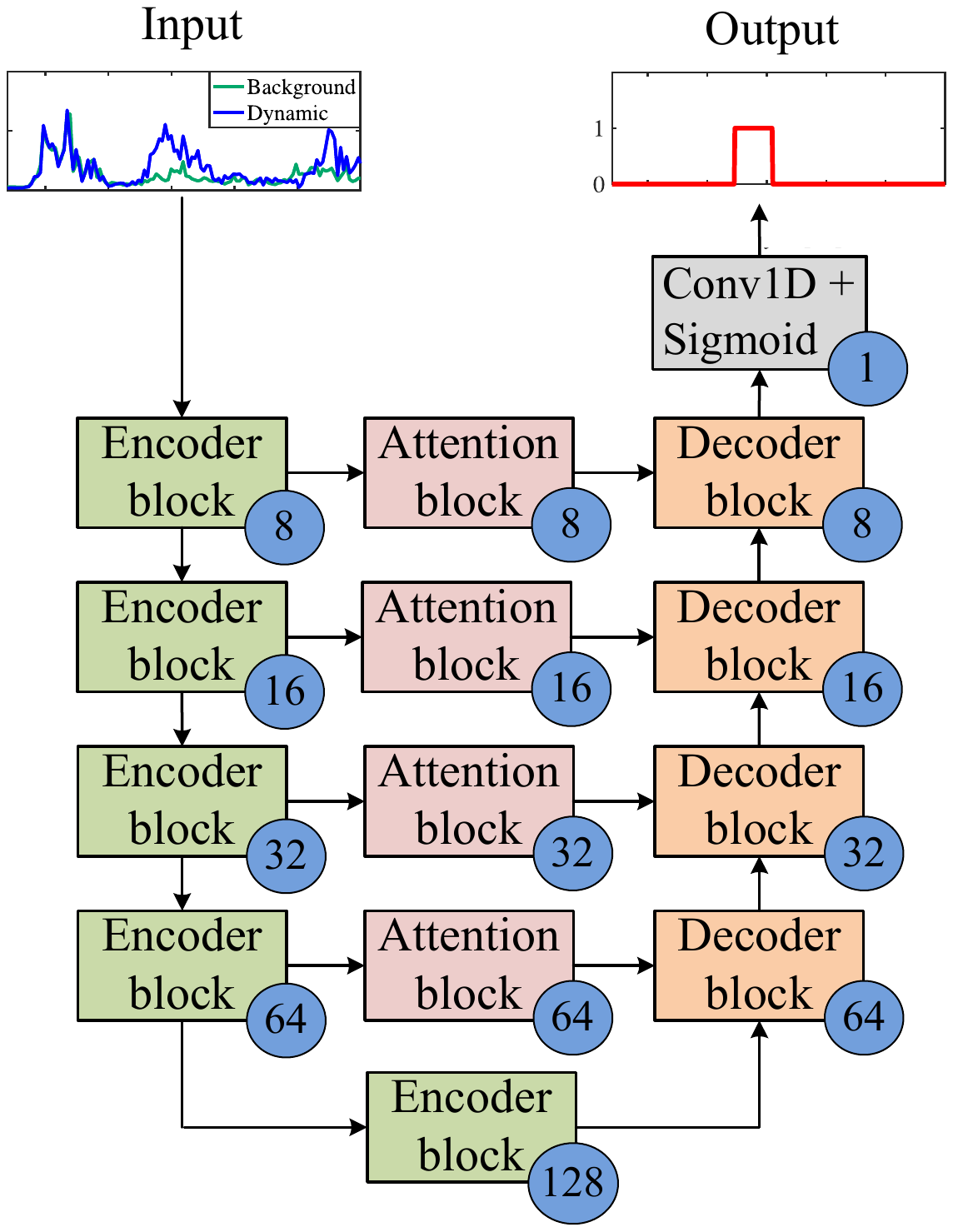}
\caption{Implemented architecture.}
\end{subfigure} %
\begin{subfigure}{.24\textwidth}
\centering
\hspace*{-0.7cm}
\vspace{3.5mm}
\includegraphics[width=6cm]{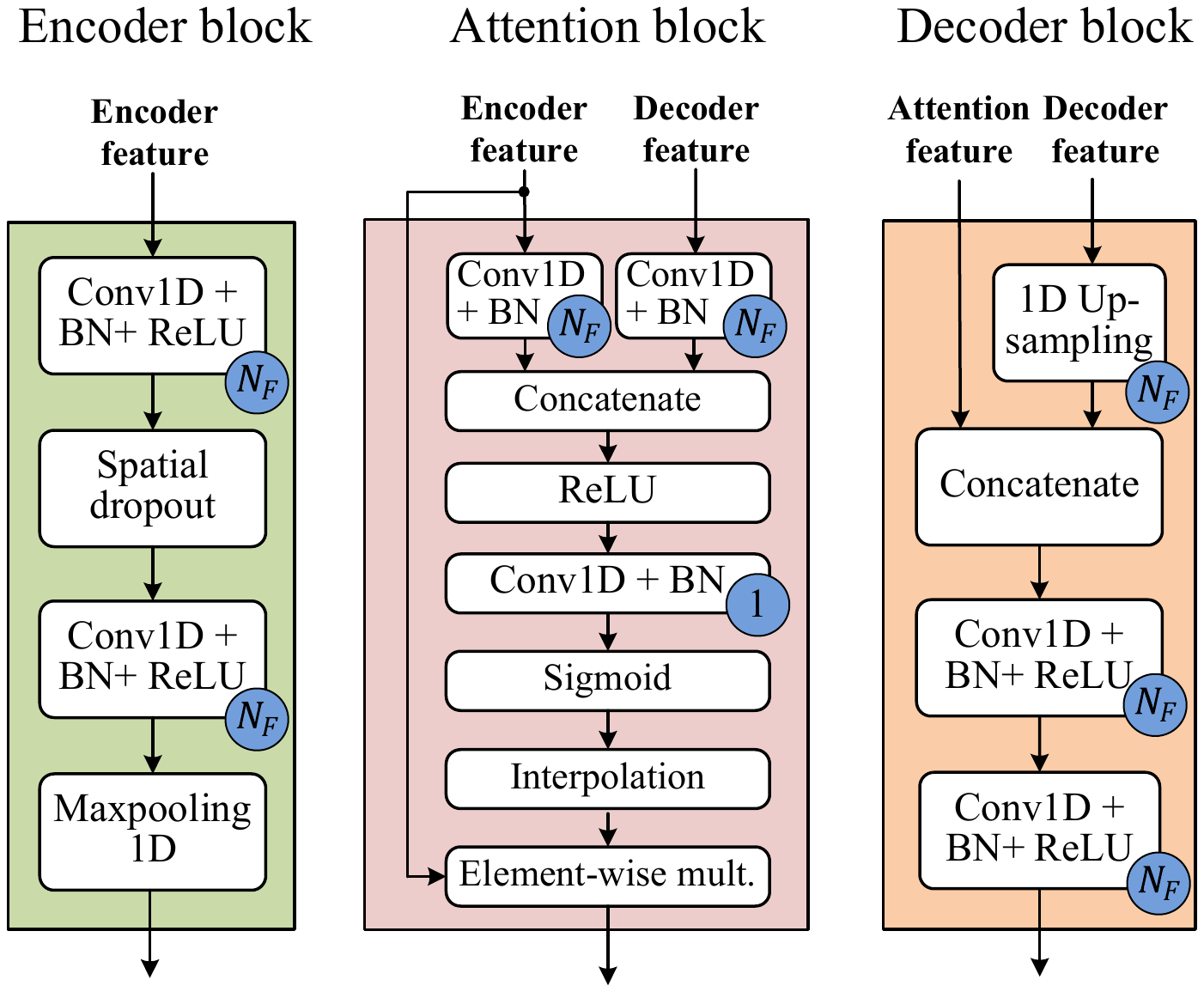}
\caption{Blocks of the 1-D attention U-Net.}\centering
\end{subfigure}
\caption{Implemented 1-D attention U-Net.}
\label{Fig4}
\vspace{-2mm}
\end{figure}

The DL-based RoI search identifies specific regions in the CIR variance profile that are likely to correspond to the target path delays caused by target-induced reflections.
This approach simplifies the task by detecting boundaries instead of estimating a single value of the target path delay via regression. 
Also, it can mitigate ambiguity from noise components by highlighting regions expected to correspond to target-induced signals.

In this paper, we use a 1-D attention U-Net for semantic segmentation to classify each pixel in a 1-D image (signal) into distinct classes. 
The network follows an encoder-decoder architecture with attention blocks in skip connections \cite{Oktay:2018}, as shown in Fig.~3(a). 
The encoders extract features and decoders generate the segmentation map (i.e., RoI). 
In addition, attention blocks enable the network to focus on important region(s) in the encoder features and ignore irrelevant ones.
Hence, it can be effective in low SNR conditions where the signal difference between the background CIR $\sigma_i^B(\tau)$ and the dynamic CIR $\sigma_{i,k}^D(\tau)$ are subtle.

The model input is the background CIR variance $\sigma_i^B(\tau)$ and dynamic CIR variance $\sigma_{i,k}^D(\tau)$, treated as a 1-D input image where each delay $\tau$ corresponds to a pixel. The size of the input is 500$\times$2.
The output is a region map $\mathcal{R}_{i,k}(\tau)$ indicating the probability of each pixel belonging to the RoI. The output dimension is 500$\times$1.
Labels are set to span a $\pm1.5$~ns region around the true target delay to guide the model in focusing on target-induced signals.

The three main blocks of the architecture are depicted in Fig.~3(b). 
First, the encoder block includes two 1-D convolutional (Conv1D) layers, each with a kernel size of 5, a stride of 1, and a number of filters $N_F$ determined by the input parameters in Fig.~3(a).
Each Conv1D layer is followed by batch normalization (BN) and rectified linear unit (ReLU) activation.
In addition, a spatial dropout with a rate of 0.3 is applied after the first Conv1D layer to mitigate overfitting. 
Finally, a 1-D max pooling with a stride of 2 is used for downsampling. 
Note that the output of each encoder block is passed to the corresponding attention block at the same level and the subsequent encoder block.

Second, the attention block takes two inputs: feature maps from the encoder and gating signals from the decoder.
These inputs are processed through Conv1D with BN to match their dimensions, with strides set to 2 and 1, respectively. 
Also, these blocks are set to a kernel size of 1 and the number of filters $N_F$.
Then, concatenation (element-wise addition) and ReLU activation are applied. 
Subsequently, a Conv1D layer with a kernel, stride, and filter size of 1, and sigmoid activation generates the attention weights.
These weights are upsampled using linear interpolation to match the dimension of the encoder feature maps.
Finally, the encoder feature maps are weighted by element-wise multiplication with the corresponding attention weights.

Each decoder block takes the feature maps from the previous decoder and the attention feature maps at the same level as inputs. 
The previous decoder feature maps are upsampled using a transposed Conv1D layer with a kernel size of 3, a stride of 2, and $N_F$ filters. 
Then, these are combined with the attention feature maps through element-wise addition. 
The combined features are processed by two Conv1D layers with BN and ReLU activation, each configured with a kernel size of 5, a stride of 1, and $N_F$ filters. 
The decoder progressively generates the segmentation map across the network levels.

The output layer uses a Conv1D with all parameters (kernel size, stride, filter size) set to 1 and sigmoid activation for pixel-wise binary classification shown in Fig.~3(a).
In our design, only 57 out of 500 samples (11.4 \%) in the CIR variance profile are labeled as the target RoI (i.e., a positive class), resulting in class imbalance.
To address this issue, the focal Tversky loss function $FT(\alpha, \beta, \gamma)$ is used for training \cite{Abraham:2019}.

\begin{equation}
FT(\alpha, \beta, \gamma) = \left(1 - T(\alpha, \beta)\right)^{1/\gamma},
\end{equation}
\begin{eqnarray}
T(\alpha, \beta) = ~~~~~~~~~~~~~~~~~~~~~~~~~~~~~~~~~~~~~~~~~~~~~~~~~~~  \nonumber \\
\frac{\sum_{j=1}^N p_j g_j + \epsilon }{\sum_{j=1}^N p_j g_j + \alpha \sum_{j=1}^N p_j (1 - g_j) + \beta \sum_{j=1}^N (1 - p_j) g_j + \epsilon},
\end{eqnarray}

where $T(\alpha, \beta)$ is the Tversky index, $j$ is the index of a pixel, $N$ is the total number of pixels, $p_j$ is the predicted probability for the $j$-th pixel after the output layer, $g_j$ is the ground truth label for the $j$-th pixel ($1$ for positive, $0$ otherwise), and $\epsilon$ is a small constant for numerical stability.  
The hyperparameters of the loss function control $\alpha$ and $\beta$ control penalties for false positives and false negatives, respectively.
In addition, the loss function prioritizes misclassified prediction results when $ \gamma > 1$.
In this study, we set $\alpha = 0.7$,  $\beta = 0.3$ , and $\gamma = 2 $.

\begin{figure}
\begin{subfigure}{.20\textwidth}
\hspace*{-0.5cm}
\centering
\includegraphics[width=4cm]{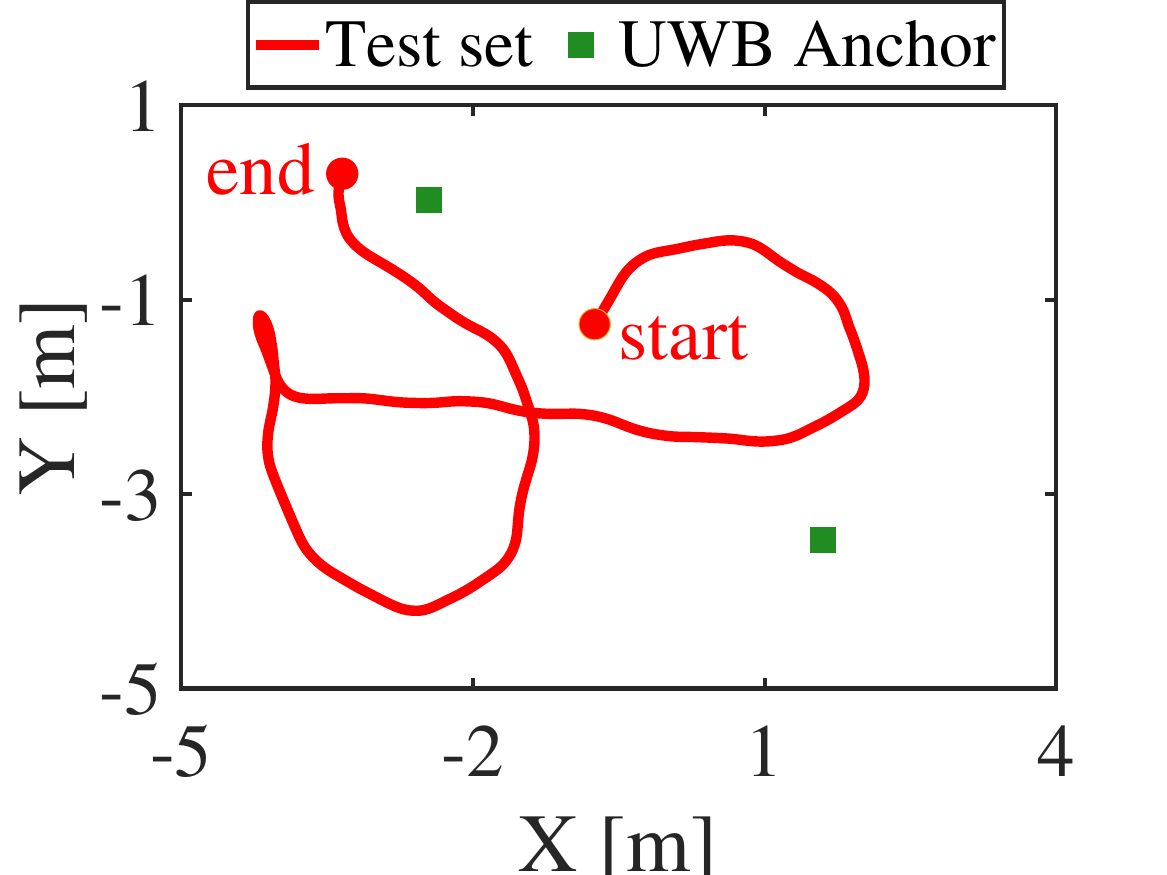}
\caption{Experimental trajectory for inference.}
\end{subfigure} %
\begin{subfigure}{.24\textwidth}
\centering
\hspace*{-0.3cm}
\vspace{3.5mm}
\includegraphics[width=5.0cm]{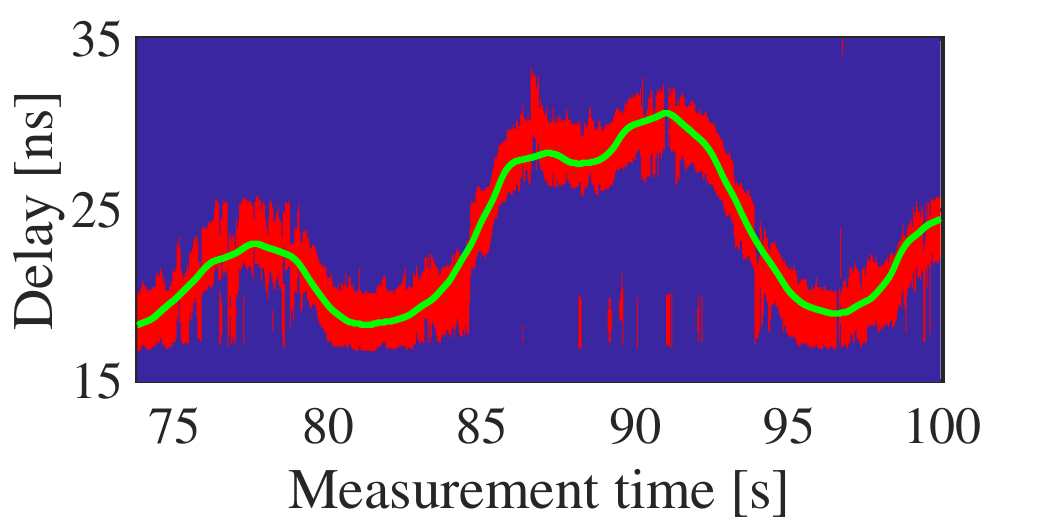}
\caption{Experimental results.}\centering
\end{subfigure}
\caption{Inference results of 1-D attention U-Net.}
\label{Fig4}
\vspace{-2mm}
\end{figure}

In the inference stage, each delay (pixel) in $\mathcal{R}_{i,k}(\tau)$ is considered part of the RoI if its probability value exceeds 0.5. 
As shown in Fig.~4(b), the predicted RoIs closely match the ground truth (green curve) and remain robust in low SNR, such as when the UWB transceivers and the target are far apart [see Fig.~4(a)]. 
Please refer to Fig.~8(b) for the training dataset for these results.

\subsection{Signal masking}

\begin{figure}
\begin{subfigure}{.24\textwidth}
\hspace*{-0.5cm}
\centering
\includegraphics[width=4.2cm]{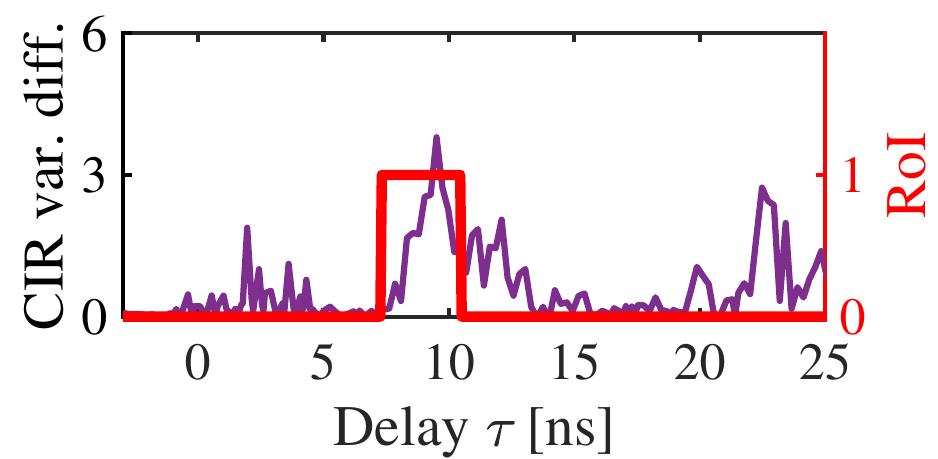}
\caption{Difference of CIR variance \newline and region map }
\end{subfigure} %
\begin{subfigure}{.24\textwidth}
\centering
\hspace*{-0.5cm}
\vspace{3.5mm}
\includegraphics[width=4.2cm]{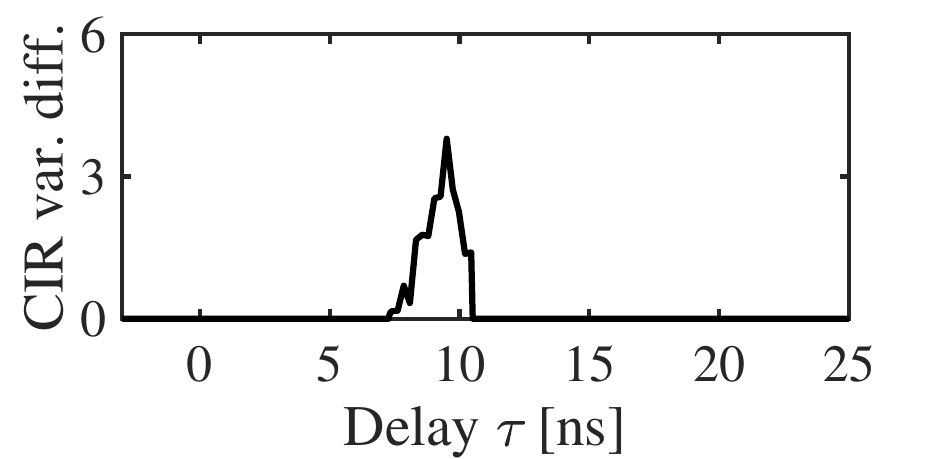}
\caption{Masked CIR variance.}
\end{subfigure}
\caption{Signal masking operation.}
\label{Fig4}
\vspace{-2mm}
\end{figure}

As shown in Fig.~5, the signal masking operation is used to suppress irrelevant noise components in the difference of CIR variance $\Delta \sigma_{i,k}(\tau)$ using the region map for the target path $\bm{\mathcal{R}}_{i,k}(\tau)$. 
This step enables only the relevant components of the difference in the CIR variance profile to be passed to the subsequent particle filter.

To be specific, for a given channel link pair $i$ and time index $k$, the masked CIR variance (with background-subtracted) $\tilde{\Delta \sigma}_{i,k}(\tau)$ is computed as:

\begin{equation}
\tilde{\Delta \sigma}_{i,k}(\tau) = \Delta \sigma_{i,k}(\tau) \odot \mathcal{R}_{i,k}(\tau), 
\label{eq:masked_cir_variance}
\end{equation}

where $\odot$ is the element-wise multiplication operator. 
In this paper, to ensure numerical stability during subsequent processing, a small constant (e.g., $10^{-12}$) is added to elements with zero value in the masked CIR variance.

\subsection{Deep Learning-assisted Particle Filter}

The final step of the proposed system feeds the multiple masked CIR variances $\tilde{\Delta \sigma}_{i,k}(\tau)$ into a particle filter to estimate the target's position. 
The key idea is utilizing DL-identified RoIs to enable the particle filter to focus on the most relevant regions within the CIR variance profiles across all channel pairs. 
In this section, we describe the implementation of the designed likelihood and the particle filter.

\subsubsection{Likelihood Calculation}

\begin{figure}
\begin{subfigure}{.24\textwidth}
\hspace*{-0.5cm}
\centering
\includegraphics[width=4.6cm]{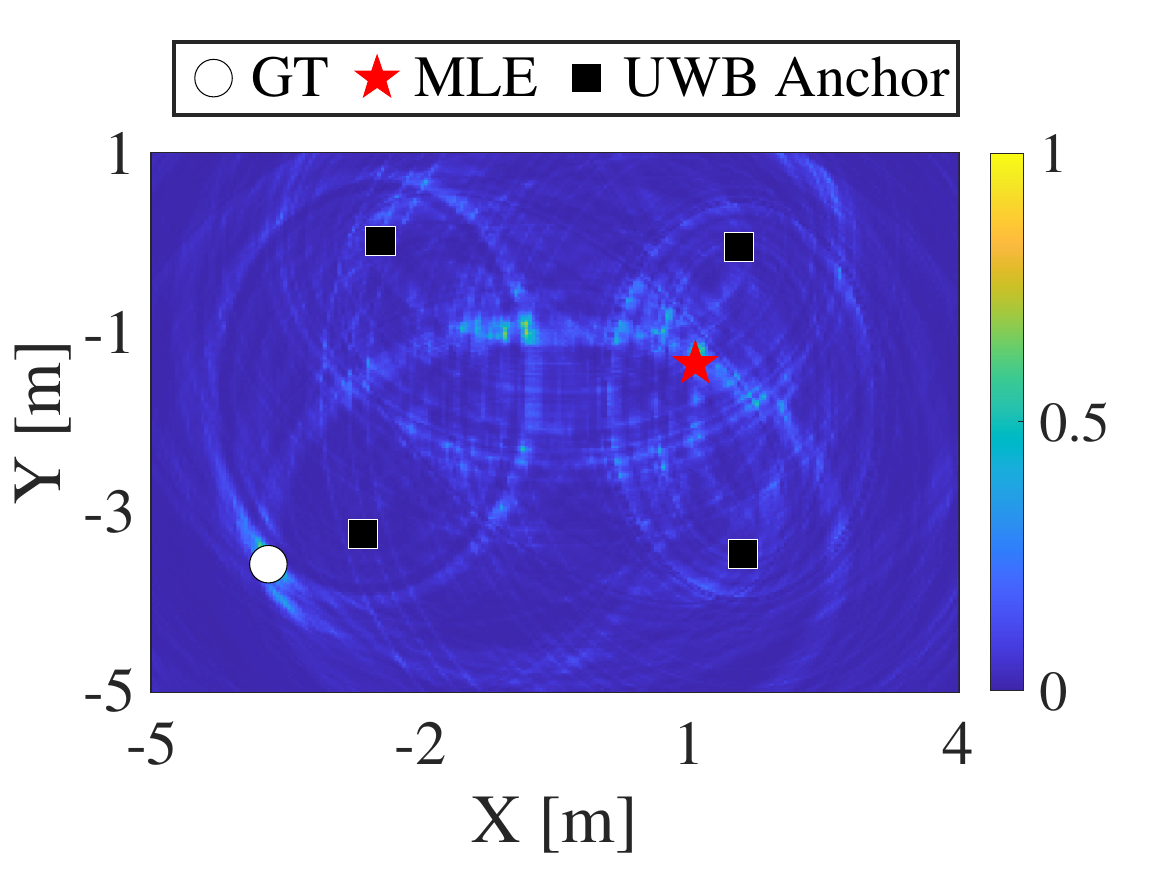}
\caption{Raw data-based MLE.}\centering
\end{subfigure} %
\begin{subfigure}{.24\textwidth}
\centering
\hspace*{-0.5cm}
\includegraphics[width=4.6cm]{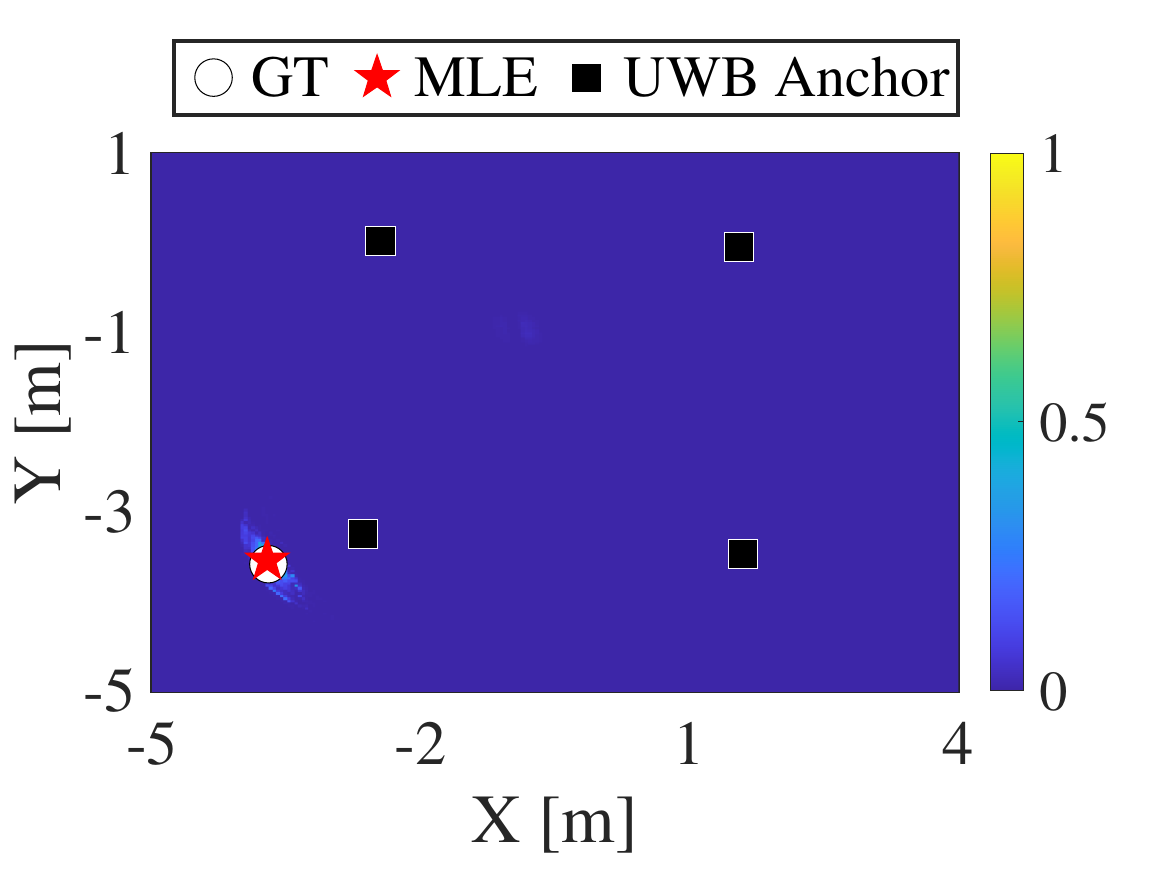}
\caption{Attention U-Net-based MLE.}\centering
\end{subfigure}
\caption{Comparison of MLE results.}
\vspace{-2mm}
\end{figure}

Before discussing the particle filter, calculating likelihood using multiple masked CIR variances $\tilde{\Delta \sigma}_{i,k}(\tau)$ is explained. 
If the target position $\mathbf{P}_k = [x_k, y_k]^T$ is known at a given time index $k$, the predicted target path delay for the $i$-th transceiver pair $\tau_{i,k}^{(\mathbf{P}_k)}$ can be computed as:

\begin{equation}
\tau_{i,k}^{(\mathbf{P}_k)} = \frac{\|\mathbf{P}_k - \mathbf{P}_{\text{Tx}}^i\| + \|\mathbf{P}_k - \mathbf{P}_{\text{Rx}}^i\|}{c}, 
\end{equation}

where $\mathbf{P}_{\text{Tx}}^i$ and $\mathbf{P}_{\text{Rx}}^i$ represent the positions of the $i$-th pair of the Tx and the Rx, $c$ is the speed of light, and $\|\cdot\|$ is the norm-2 operator.

To calculate the likelihood, the masked CIR variance $\tilde{\Delta \sigma}_{i,k}(\tau)$ is normalized such that its integral equals 1, allowing it to be treated as a probability density function (PDF)~\cite{Mazuelas:2018}. 
The PDF of $\tilde{\Delta \sigma}_{i,k}(\tau)$ is defined as:

\begin{equation}
p(\tilde{\Delta \sigma}_{i,k} \mid \tau) \triangleq \frac{\tilde{\Delta \sigma}_{i,k}(\tau)}{\int \tilde{\Delta \sigma}_{i,k}(\tau)  \mathrm{d} \tau}. 
\end{equation}

Then, for each channel link $i$, the likelihood $p(\tilde{\Delta \sigma}_{i,k} \mid \tau = \tau_{i,k}^{(\mathbf{P}_k)})$ is evaluated at the predicted target path delay $\tau_{i,k}^{(\mathbf{P}_k)}$. 
Finally, multiplying the individual likelihoods across all channel pairs, the position-related joint likelihood function $\mathcal{L}(\mathbf{P}_k)$ at a specific target position $\mathbf{P}_k$ can be defined as: \footnote{
Since the measurements are obtained from different channel links, they are independent of each other (i.e., uncorrelated).}

\begin{equation}
\mathcal{L}(\mathbf{P}_k) \triangleq \prod_{i=1}^{N_\text{UWB}^\text{Pair}} p(\tilde{\Delta \sigma}_{i,k} \mid \tau = \tau_{i,k}^{(\mathbf{P}_k)}),
\end{equation}

where $N_\text{UWB}^\text{Pair}$ is the total number of channel link pairs.

The joint likelihood distributions $\mathcal{L}(\mathbf{P}_k)$ over all possible positions are shown in Fig.~6(a) and (b).
Fig.~6(a) shows the maximum likelihood estimation (MLE) result obtained using CIR variance profiles without signal masking.
We can observe that estimating position becomes difficult because multiple modes appear. 
On the other hand, the proposed method generates a single-mode distribution and this MLE result is closer to the ground truth (GT) shown in Fig.~6(b). 

Although this joint likelihood distribution could be used to directly estimate the position via MLE, evaluating $\mathcal{L}(\mathbf{P}_k)$ for all possible positions is computationally expensive. 
Hence, we consider a particle filter that uses a finite set $M$ of particles (e.g., $M = 200$) while maintaining localization performance.

\subsubsection{Particle Filter Implementation}

\begin{figure}
\includegraphics[width=5cm]{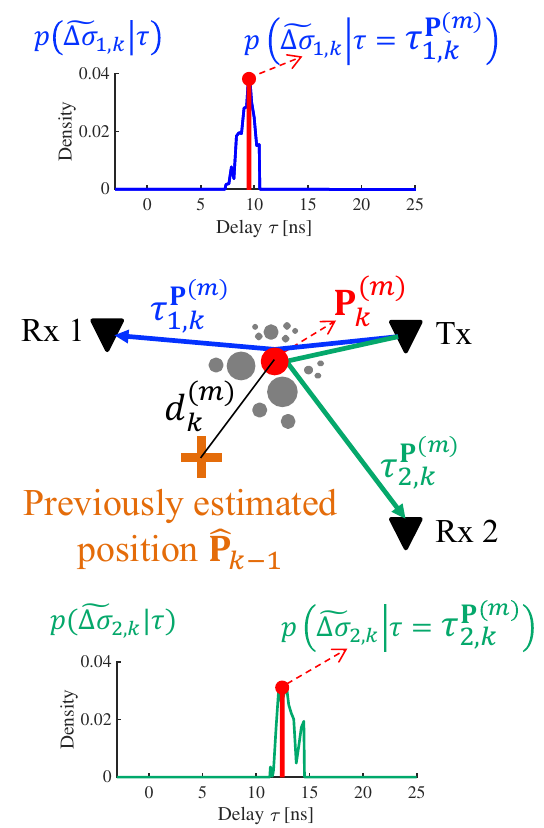}\centering
\caption{The update step of the particle filter.}
\end{figure}

The particle filter recursively estimates the posterior distribution of the target's position using the likelihood and the prior distribution derived from the target's motion dynamics. 
This algorithm uses a finite set of particles to approximate the posterior distribution. 
In this paper, the state to be estimated is the position and velocity of the target in the 2-D Cartesian coordinates, $\mathbf{x}_k = [x_k \ y_k \ \mathrm{v}^x_k \  \mathrm{v}^y_k]^T$. 
The implemented particle filter consists of three main steps.

\paragraph{Prediction step}  
All particles are propagated using a motion dynamics model that predicts the potential target movement based on the state of the particles at the previous time. 
In this paper, we use a nearly constant velocity (CV) model.  
Also, Gaussian random noise with a standard deviation of  $\sigma_{cv}$ is considered for the uncertainty of the target's movement. 
Following the hyperparameter in \cite{Li:2022:MAP}, we set it to 0.3.

\paragraph{Update step}  
This step assigns higher weights to particles that are consistent with both the observed data and the previously estimated position, as shown in Fig.~7.
The weight of the $m$-th particle is updated based on the joint likelihood $\mathcal{L}(\mathbf{P}_k^{(m)})$ and its proximity to the previously estimated position $\hat{\mathbf{P}}_{k-1}$, where $\mathbf{P}_k^{(m)}$ is the $m$-th particle’s position. The weight of the $m$-th particle $w^{(m)}$ is calculated as:

\begin{equation}
w^{(m)} = \exp\left(-\frac{(1 - \tilde{\mathcal{L}}(\mathbf{P}_k^{(m)}))^2}{2\sigma_L^2}\right) \exp\left(-\frac{\|\hat{\mathbf{P}}_{k-1} - \mathbf{P}_k^{(m)}\|^2}{2\sigma_d^2}\right),
\end{equation}

where $\tilde{\mathcal{L}}(\mathbf{P}_k^{(m)})$ represents the normalized joint likelihood for the $M$ particles to $(0, 1]$, and the hyperparameters $\sigma_L$ and $\sigma_d$ control the sensitivity to likelihood and spatial proximity, respectively. 
In this paper, these are set to 0.5 and 1.

According to the dataset~\cite{Ledergerber:2020}, the UWB CIR measurement rate for a channel pair is about 200~Hz. 
Hence, a pedestrian is expected to move within a centimeter between measurements. 
For this reason, it is reasonable to consider spatial proximity. 

Unlike the grid-based MLE method used to evaluate the difference from the global maximal point as in~\cite{Li:2022:MAP}, the proposed method evaluates particles locally within the RoIs. 
This approach significantly reduces computational cost while maintaining performance.

\paragraph{Resampling step}  
To mitigate the degeneracy problem of the particle filter, some particles are resampled based on their weights. 
Particles with high weights are duplicated, and particles with low weights are discarded. 
This process redistributes the particle set to focus on areas with higher probability density. 
In this paper, systematic resampling is used~\cite{Arulampalam:2002}.

This procedure is repeated for each incoming CIR measurement from all channel link pairs, and the estimated position at time $k$  $\hat{\mathbf{P}}_k$ is obtained by calculating the weighted average of the particles.

\section{Evaluation}
\subsection{Experimental Data}

\begin{figure}
\centering
\begin{subfigure}{.24\textwidth}
\hspace*{-0.5cm}
\includegraphics[width=4.6cm]{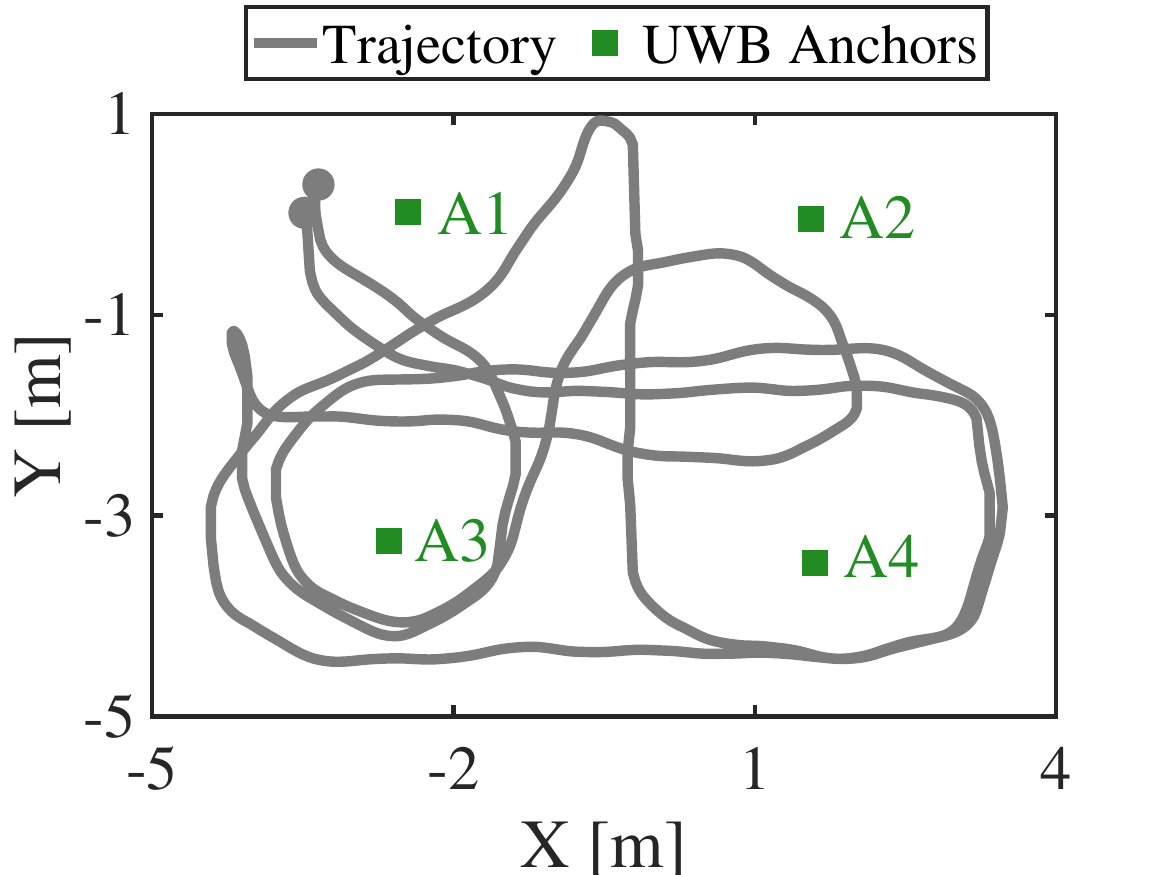}
\caption{Experimental trajectory.}\centering
\end{subfigure} %
\begin{subfigure}{.24\textwidth}
\hspace*{-0.5cm}
\includegraphics[width=4.6cm]{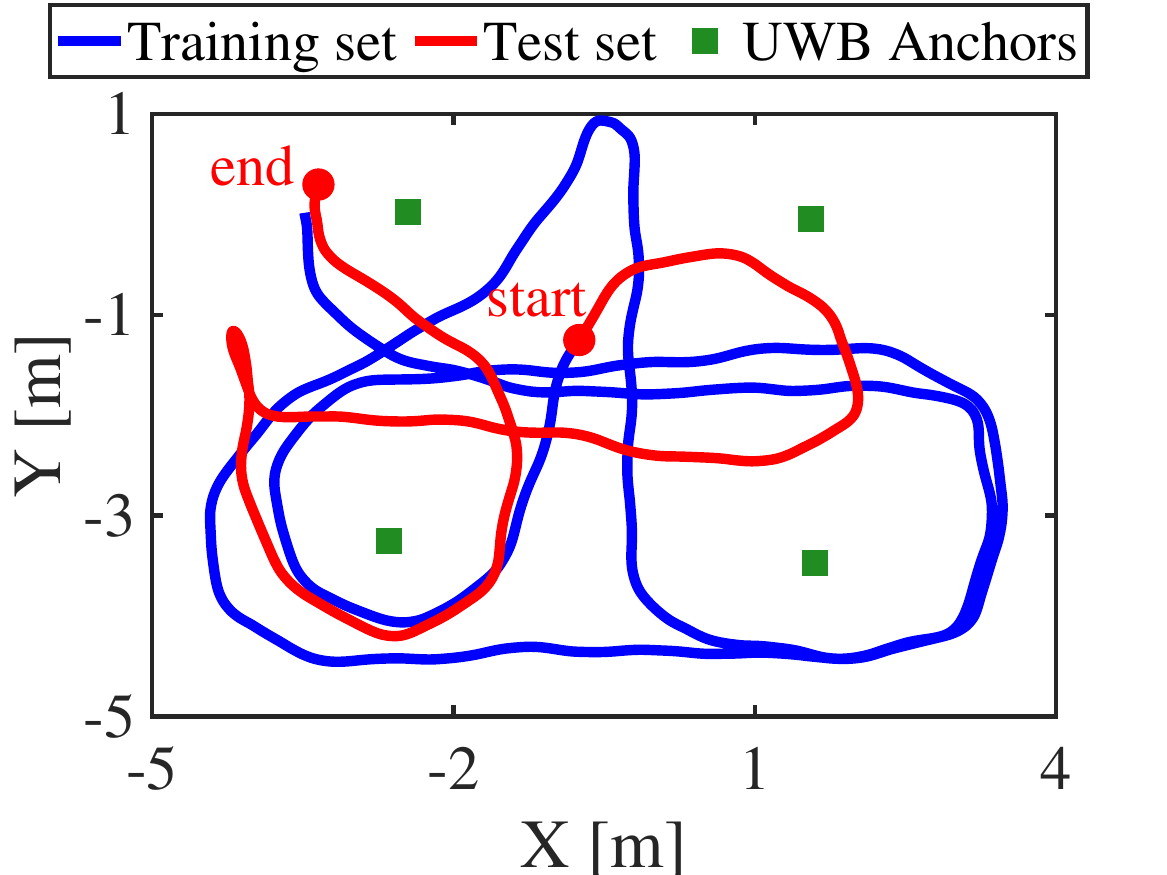}
\caption{CASE-1.}\centering
\end{subfigure} %
\vspace{2mm}
\begin{subfigure}{.24\textwidth}
\vspace{2mm}
\hspace*{-0.5cm}
\includegraphics[width=4.6cm]{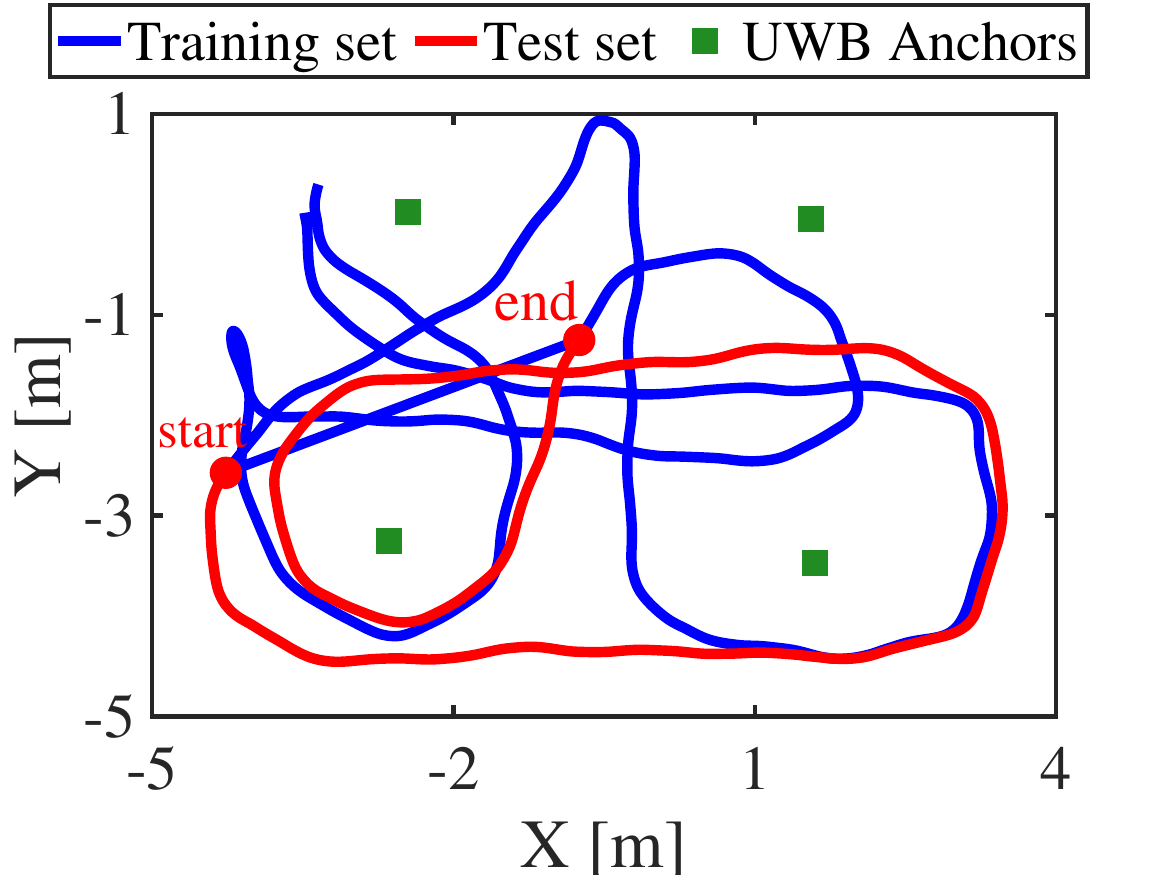}
\caption{CASE-2.}\centering
\end{subfigure} %
\begin{subfigure}{.24\textwidth}
\hspace*{-0.5cm}
\includegraphics[width=4.6cm]{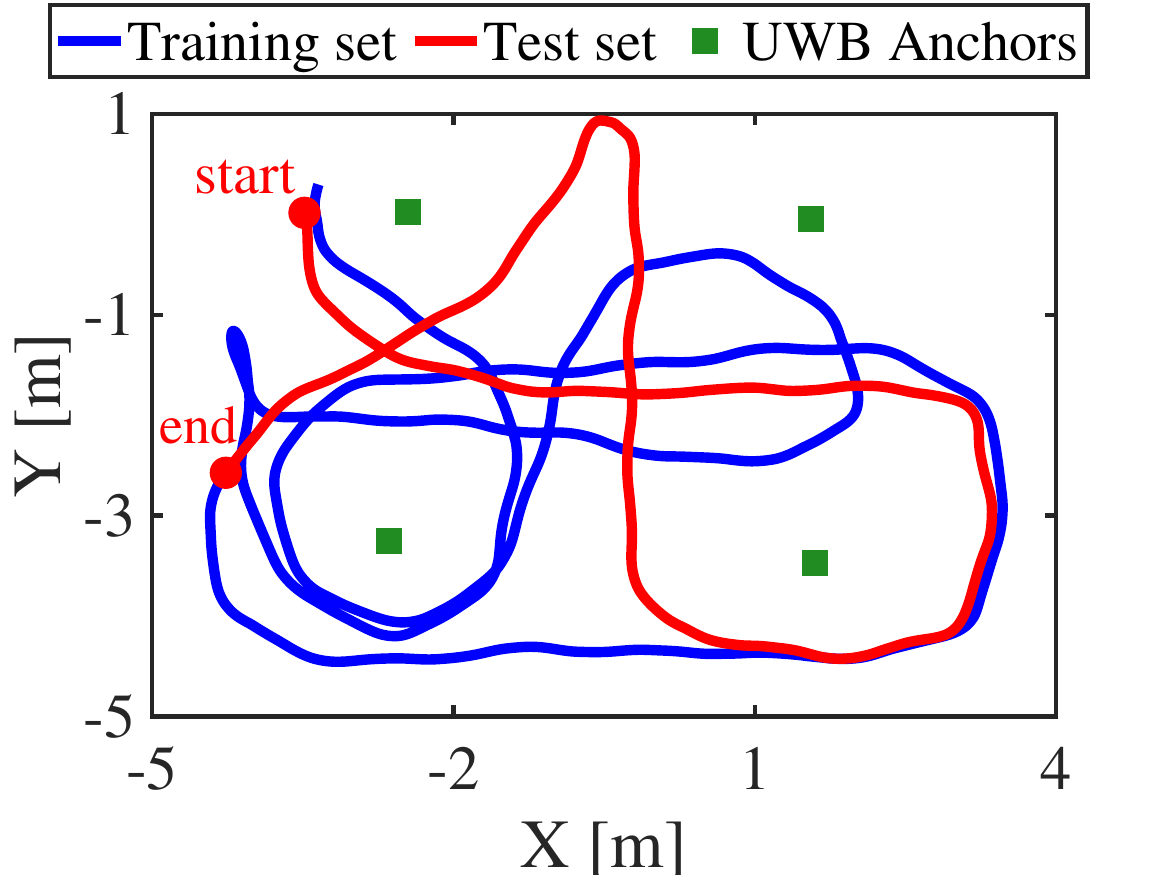}
\caption{CASE-3.}\centering
\end{subfigure}
\caption{Experimental environment.}
\vspace{-2mm}
\end{figure}

As shown in Fig~8(a), the evaluation is performed using a public dataset with four COTS UWB devices (Qorvo’s DWM1000), deployed in the multistatic radar setup~\cite{Ledergerber:2020}.
Note that all UWB transceivers operate as Tx and Rx simultaneously. 
Therefore, there are a total of six pairs of channel links (i.e., $N_\text{UWB}^\text{Pair} = 6$)
\footnote{
We assume channel reciprocity between transceivers. It reduces the number of unique channel pairs from 12 to 6, decreasing the computational load of the DL model.}.
The UWB transceivers operated at a carrier frequency of 3993.6~MHz (Channel 4) and a pulse repetition frequency of 16~MHz. 
Each anchor communicated with the others, measuring CIR at an average rate of 188~Hz. 
The ground truth of the experimenter’s trajectory was obtained via a motion capture system. 
For details on the experimental procedures, please refer to~\cite{Ledergerber:2020}.

\subsection{Experimental Results using four anchors}

\footnotesize
\begin{table}[t]
\caption{Evaluation for DL models averaged over all cases.}
\vspace{-2mm}
\begin{center}
\begin{tabular}{|c|c|c|c|c|}
\hline
\textbf{DL Model} & \textbf{Recall} & \textbf{Precision} & \textbf{F1 Score} & \textbf{Inference speed} \\
\hline
\hline
Attention U-Net &0.865 & 0.776 & 0.818 & 0.573~ms \\
\hline
Vanilla U-Net   & 0.820 & 0.797 & 0.808 & 0.552~ms \\
\hline
\end{tabular}
\end{center}
\vspace{-3mm}
\label{tab:unets_evalution_four_anchors}
\end{table}
\normalsize

In this section, we evaluate the performance of the proposed system using all available UWB anchors.
Since the proposed method utilizes the DL model, we divide the dataset into a training set and a test set. 
According to~\cite{Li:2022:CNN}, the measured time of an experimenter's movement ranges from 21.4 seconds to 100 seconds (i.e., 78.6 seconds). 
The training set consists of 52.4 seconds (66.7\% of the total trajectories) and the test set of 26.2 seconds.
Fig.~8(b)-(d) shows three cases of dataset splits.

We first discuss the performance of the DL-based RoI search.
In this section, the Attention U-Net and the vanilla U-Net \cite{Ronneberger:2015} are evaluated. 
The vanilla U-Net is implemented the same as the attention U-Net except that it excludes the attention blocks from the skip connections.
For training and inference, we use TensorFlow 2.13.0 with Python 3.8.15 on an Apple M1 Pro laptop. 
The optimizer used for training is Adam with a learning rate of 0.001. 
The training was conducted with a batch size of 128 and 10 epochs.

The metrics used to evaluate the U-Net models are recall, precision, and F1 Score.
Recall measures the proportion of correctly predicted positive cases (i.e., target RoIs) among all true positives. 
And, precision represents the proportion of correctly predicted positive cases among all predicted as positive. 
Finally, the F1 Score is calculated as the harmonic mean of precision and recall, also known as the dice score.
This is one of the metrics commonly used to evaluate semantic segmentation models.

The performance of the U-Net models is summarized in Table I. 
Each metric is averaged across CASE-1, CASE-2, and CASE-3 to provide a comprehensive assessment.
The recall of the Attention U-net is 0.045 higher than the vanilla U-net. 
Note that the recall is considered the most important metric because it reflects the model's ability to accurately identify all positive cases (i.e., target RoIs).
Also, since we combine UWB CIR measurements to perform localization, the slight degradation of the model's precision is negligible.
Finally, the inference speed between the two models is almost the same.

\begin{figure}
\centering
\begin{subfigure}{.23\textwidth}
\hspace*{-0.3cm}
\includegraphics[width=4.6cm]{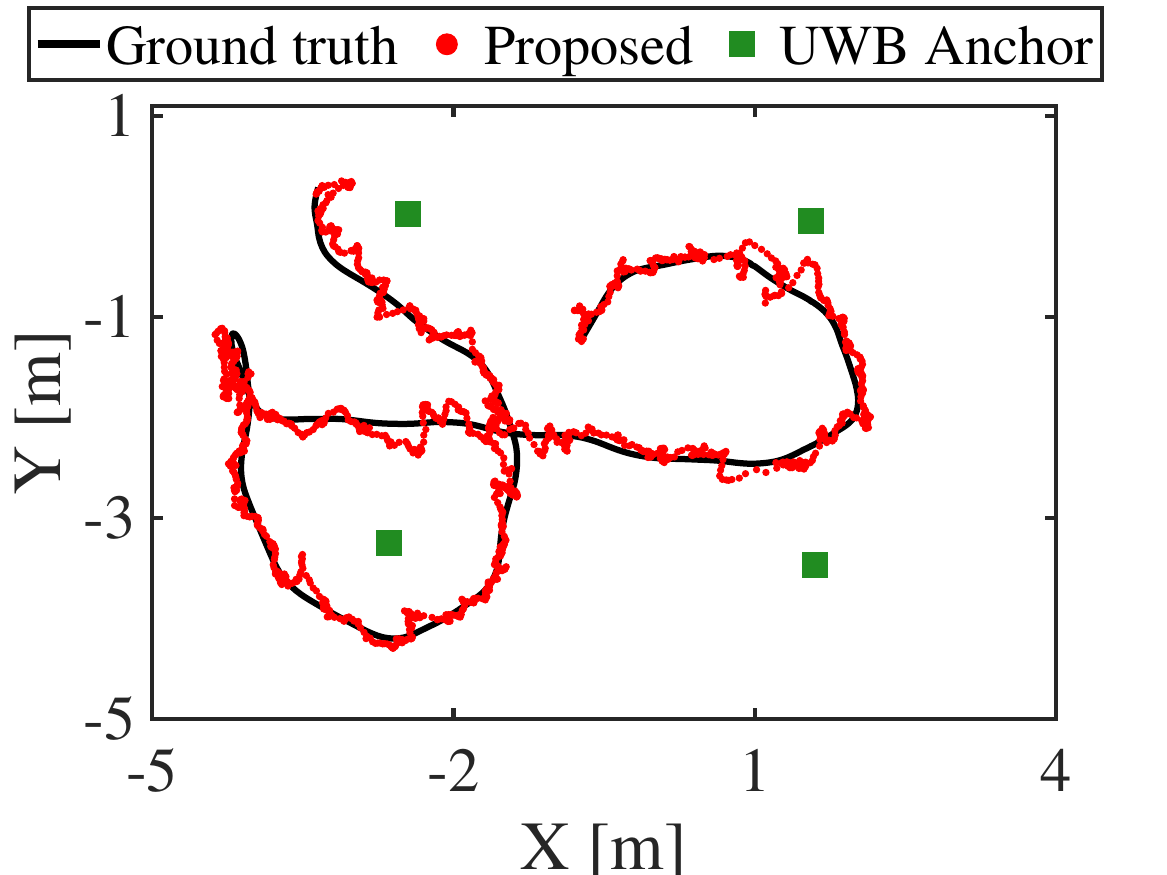}
\caption{CASE-1 result.}\centering
\end{subfigure} %
\begin{subfigure}{.23\textwidth}
\hspace*{-0.0cm}
\includegraphics[width=4.6cm]{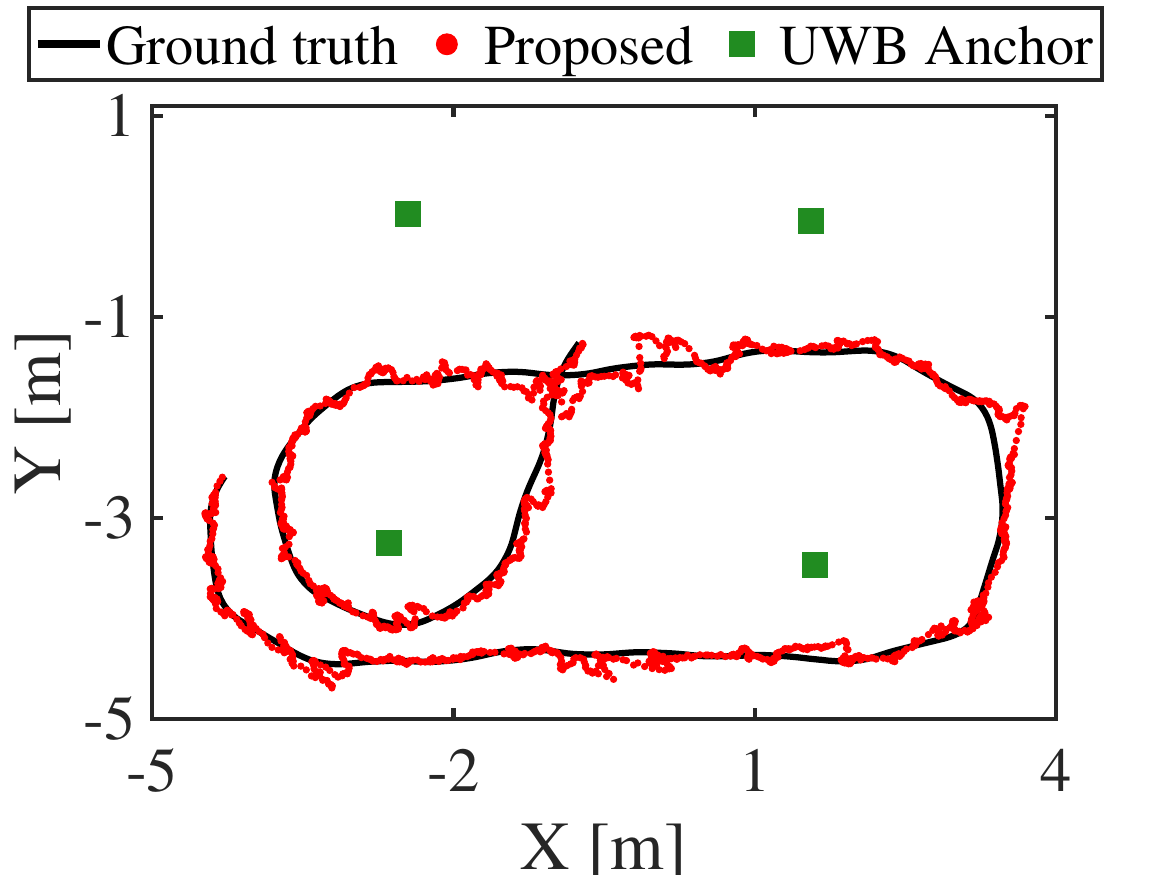}
\caption{CASE-2 result.}\centering
\end{subfigure} %
\vspace{2mm}
\begin{subfigure}{.23\textwidth}
\vspace{2mm}
\hspace*{-0.3cm}
\includegraphics[width=4.6cm]{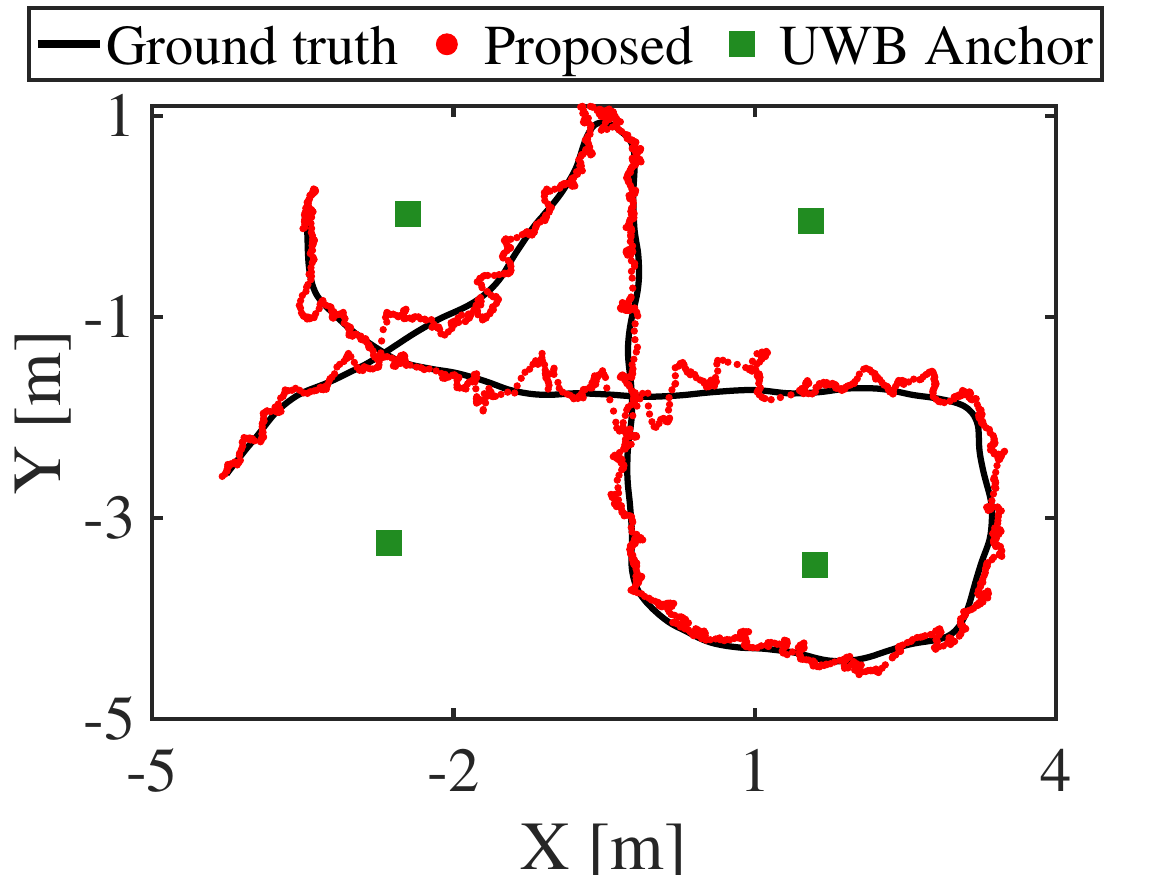}
\caption{CASE-3 result.}\centering
\end{subfigure} %
\begin{subfigure}{.23\textwidth}
\hspace*{-0.0cm}
\includegraphics[width=4.6cm]{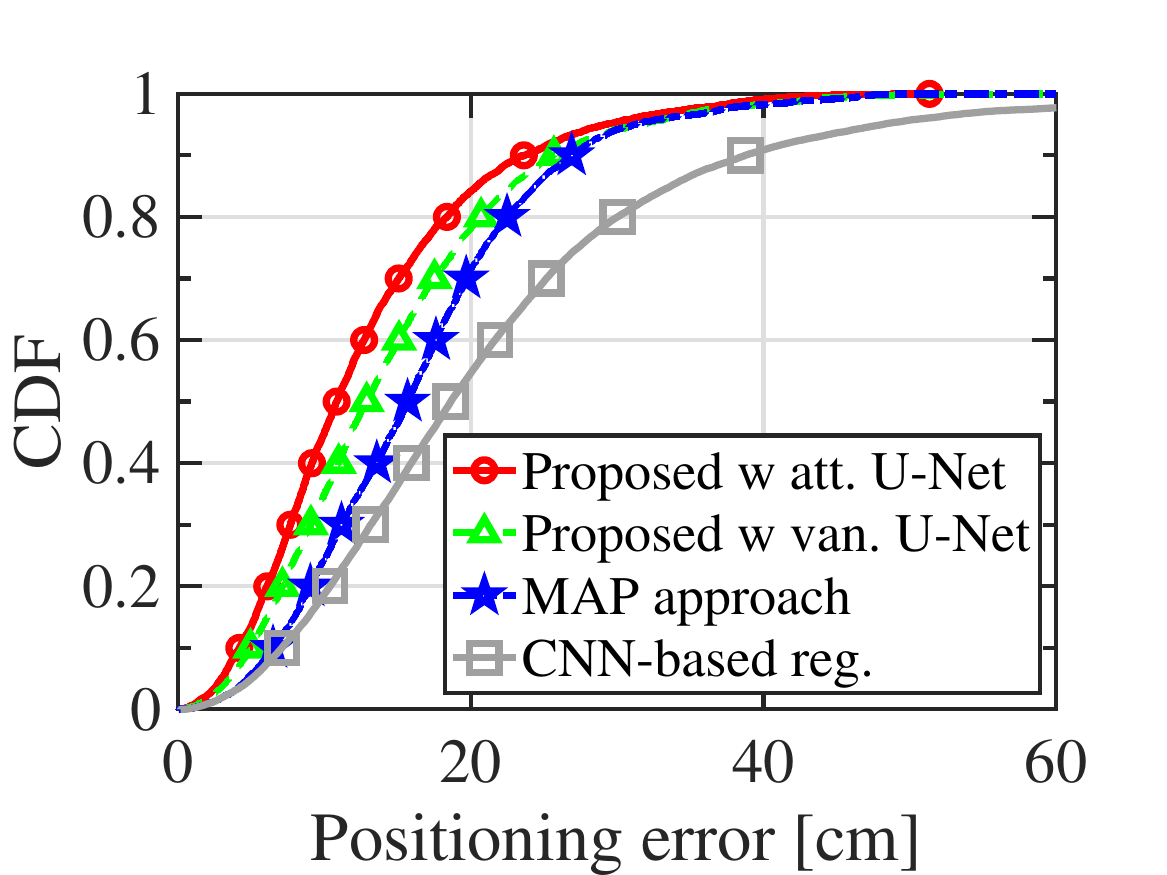}
\caption{CDF for comparison.}\centering
\end{subfigure}
\caption{Experimental results with four anchors.}
\vspace{-2mm}
\end{figure}

\footnotesize
\begin{table}[t] 
\caption{Tracking accuracy across the experimental cases.}
\vspace{-2mm}
\begin{center}
\begin{tabular}{|c|c|c|c|}
\hline
\textbf{Case} & \textbf{2-D RMSE} & \textbf{50th Error} & \textbf{90th Error} \\ 
\hline
\hline
CASE-1 & 15.3 cm & 10.7 cm & 23.7 cm \\ 
\hline
CASE-2 & 14.0 cm & 10.4 cm & 21.6 cm \\ 
\hline
CASE-3 & 15.9 cm& 11.8 cm & 25.3 cm\\ 
\hline
\end{tabular}
\end{center}
\vspace{-3mm}
\end{table}
\normalsize

Then, the proposed method (namely, proposed with attention U-Net) is evaluated across three cases, and the results are measured using 2-D root mean square error (RMSE), 50th percentile, and 90th percentile errors. 
Table~II summarizes the tracking accuracy for all cases. 
The proposed method achieves accurate tracking performance over all cases with a 2-D RMSE of  15.1~cm, a 50th percentile error of 10.8~cm, and a 90th percentile error of 23.6~cm. 
The visualization of the localization results is presented in Fig.~9(a)-(c).

\footnotesize
\begin{table}[t]
\caption{Comparison of tracking accuracy using four anchors with existing methods.}
\vspace{-2mm}
\begin{center}
\begin{tabular}{|p{2.35cm}|c|c|c|}
\hline
\textbf{Methods} & \textbf{50th Error} & \textbf{90th Error} & \textbf{DL}\\ 
\hline
\hline
CNN-based reg. & 18.5~cm & 38.8~cm & O\\ 
\hline
MAP approach & 15.6~cm & 27.2~cm & X\\ 
\hline
Proposed method \newline with vanilla U-Net  & 12.9~cm & 25.7~cm & O\\ 
\hline
Proposed method \newline with attention U-Net & 10.8~cm & 23.6~cm & O\\ 
\hline
\end{tabular}
\end{center}
\vspace{-3mm}
\label{tab:comparison_four_anchors}
\end{table}
\normalsize

The proposed method is compared to our variant (namely, proposed with vanilla U-Net) and several state-of-the-art (SOTA) approaches, as summarized in Table~III. 
Our method reduces the 50th percentile error by 30.8\% compared to the MAP approach (from 15.6 cm to 10.8 cm) and by 41.6\% compared to CNN regression with CIR variance (from 18.5 cm to 10.8 cm). 
Also, for the 90th percentile error, the proposed method achieves a 13.2\% improvement over the MAP approach (from 27.2 cm to 23.6 cm) and a 39.2\% improvement over CNN regression with CIR variance (from 38.8 cm to 23.6 cm). 
It also shows better localization performance compared to the method using vanilla U-Net. This is because of the higher recall metric of the attention U-Net in Table I.
The cumulative distribution function (CDF) of positioning error for the entire range is shown in Fig.~9(d), where the proposed method demonstrates the best performance.

\footnotesize
\begin{table}[t]
\centering
\caption{Computation time and 2-D RMSE comparison across methods.}
\vspace{-2mm}
\begin{center}
\begin{tabular}{|p{2.35cm}|c|c|c|}
\hline
\textbf{Methods} & \textbf{Exec. time} & \textbf{2-D RMSE} & \textbf{DL}\\ 
\hline
\hline
CNN-based reg.                & 2.7 ms            & 25.6   cm            & O           \\
\hline
MAP approach.                 & 37.2 ms             & 18.4 cm                & X           \\
\hline
Proposed method \newline with vanilla U-Net              & 4.0 ms             & 16.9 cm                & O           \\ 
\hline
Proposed method \newline with attention U-Net              & 4.1 ms             & 15.1 cm                & O           \\ 
\hline
\end{tabular}
\end{center}
\vspace{-3mm}
\label{tab:comparison_four_anchors}
\end{table}
\normalsize

We also evaluate the proposed method on an Apple M1 Pro laptop using MATLAB R2023b to assess computation times. In terms of computational cost, the proposed method (4.1~ms) is more than about 9.1 times faster than the MAP approach (37.2~ms) with higher accuracy.
Although it is slightly slower than CNN regression with CIR variance (2.7~ms), the proposed method achieves a 41.0\% lower 2-D RMSE, representing a reasonable trade-off for improved accuracy.

\subsection{Experimental Results using three anchors}

\footnotesize
\begin{table}[t]
\caption{Evaluation of DL models.}
\vspace{-2mm}
\begin{center}
\begin{tabular}{|c|c|c|c|}
\hline
\textbf{DL Model} & \textbf{Recall} & \textbf{Precision} & \textbf{F1 Score} \\
\hline
\hline
Attention U-Net & 0.808 & 0.768 & 0.787 \\
\hline
Vanilla U-Net & 0.718 & 0.695 & 0.706 \\
\hline
\end{tabular}
\end{center}
\vspace{-3mm}
\label{tab:unets_evalution_three_anchors}
\end{table}
\normalsize

\footnotesize
\begin{table}[t]
\caption{Comparison of tracking accuracy using three anchors with existing methods.}
\vspace{-2mm}
\begin{center}
\begin{tabular}{|p{2.35cm}|c|c|c|c|}
\hline
\textbf{Methods} & \textbf{50th Error} & \textbf{90th Error} & \textbf{2-D RMSE} & \textbf{DL}\\ 
\hline
\hline
CNN-based reg. & 32.2~cm & 199.3~cm &  102.8~cm & O\\ 
\hline
MAP approach & 22.7~cm & 50.8~cm  & 35.3~cm & X\\ 
\hline
Proposed method \newline with vanilla U-Net & 19.5~cm & 53.1~cm & 31.7~cm & O\\ 
\hline
Proposed method \newline with attention U-Net & 16.5~cm & 40.8~cm & 25.5~cm & O\\ 
\hline
\end{tabular}
\end{center}
\vspace{-3mm}
\label{tab:comparison_four_anchors}
\end{table}
\normalsize

\begin{figure}
\begin{subfigure}{.24\textwidth}
\hspace*{-0.5cm}
\centering
\includegraphics[width=4.6cm]{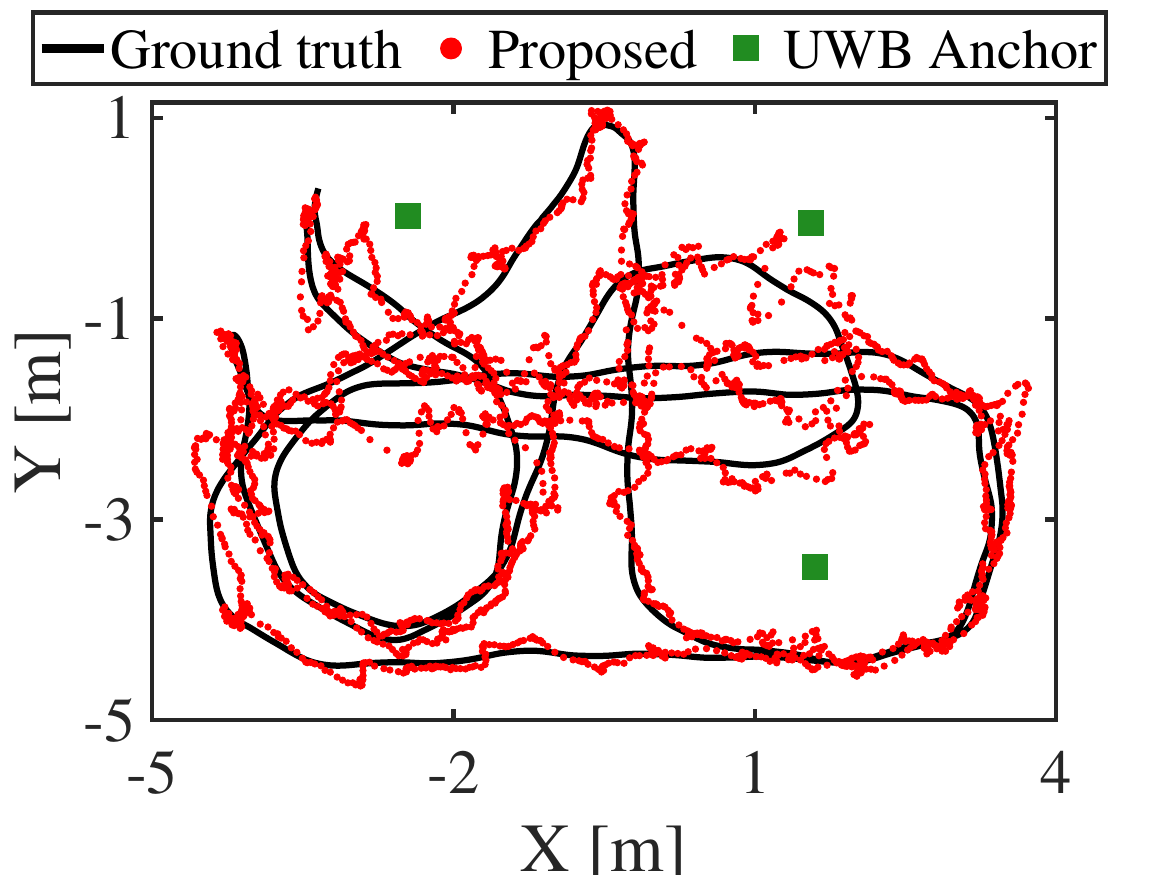}
\caption{Localization result}\centering
\end{subfigure} %
\begin{subfigure}{.24\textwidth}
\centering
\hspace*{-0.5cm}
\includegraphics[width=4.6cm]{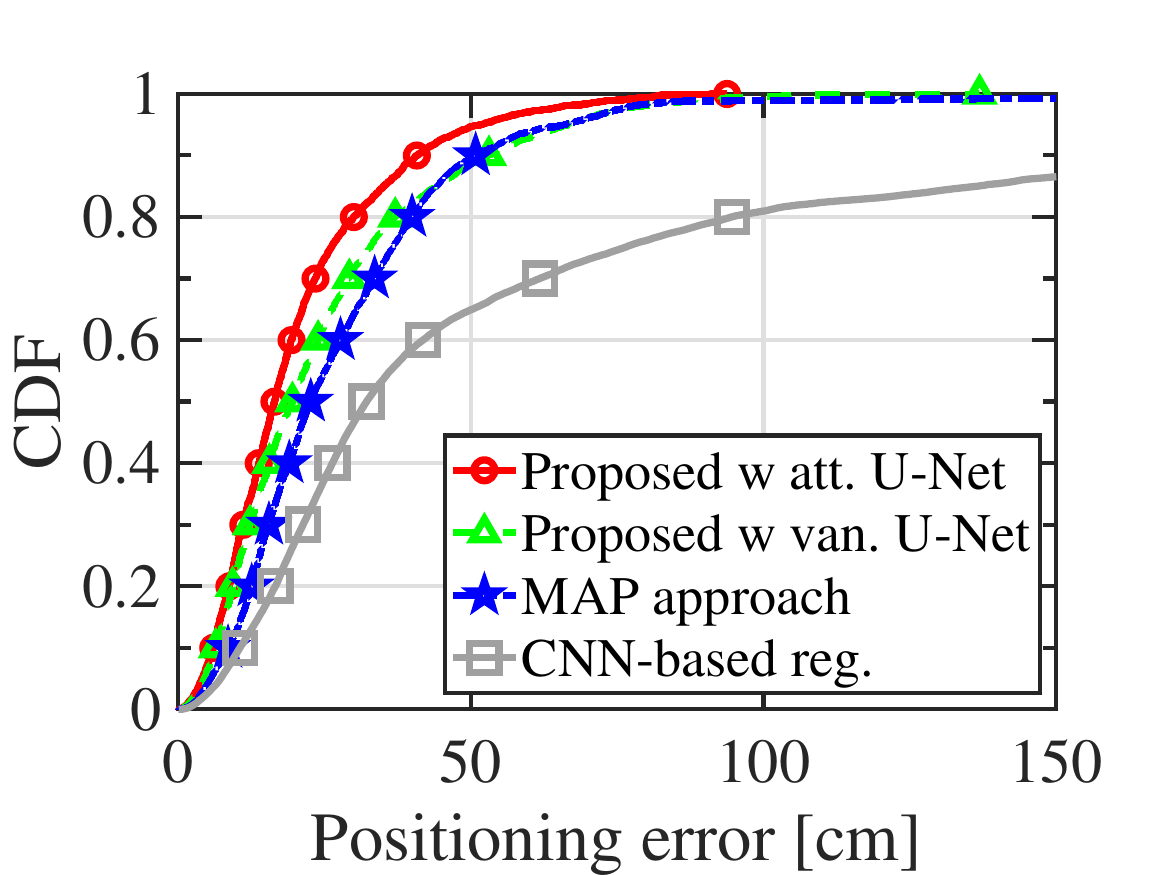}
\caption{CDF for comparison.}\centering
\end{subfigure}
\caption{Localization results with three anchors.}
\vspace{-2mm}
\end{figure}

In this section, we evaluate the proposed system's performance under the constraint of a limited number of anchors, which reflects real-world scenarios. 
The goal is to demonstrate the system's robustness and reliability while validating the generalization capability of the DL model.

For vehicle theft detection, UWB anchors are typically installed inside the vehicle and near its bumpers. 
Unfortunately, depending on the thief's position, significant signal quality degradation occurs due to obstruction caused by the vehicle's structure.
This makes precise localization challenging.
Similarly, in smart home IoT scenarios, some anchor malfunctions can be complicated localization.

This setup also allows us to validate the generalization capability of the DL model. 
According to \cite{Li:2022:CNN}, the CIR presents distinct channel characteristics for each Tx-Rx link.
These differences arise from propagation channels and environmental dependencies, such as reflections or scattering.
We also extracted the relative target path delay based on the first path assumed to be the LOS path from the measured CIR.
Moreover, the distances between the anchors used in the experiments are slightly different, so the channel environment is inherently different.
For this reason, the robustness of the DL model can be greatly reduced if data from all channel links is not included in the training dataset.

For this evaluation, we use three anchors (A1, A2, and A4) for localization purposes [see Fig. 8(a)]. So, $N_\text{UWB}^\text{Pair} = 3$. 
In addition, the other channel links associated with A3 (i.e., A1-A3, A2-A3, and A3-A4 pairs) are used as the training dataset.  
So, the datasets used for training and localization (testing) are completely separated.

First, let us evaluate the classification performance of the U-Net models. 
The training method for the models is the same as in the previous experiment.
Table V shows the classification performance of the two DL models. 
Both models experience performance degradation, but the vanilla U-Net is significantly worse by more than 0.1 on all metrics compared to Table I. 
This decrease may be due to the na\"ive model's inability to generalize, as it relies on data from specific channel pairs during the training.

In contrast, the Attention U-Net shows minimal degradation in precision and its recall decreased from 0.865 to 0.808 when evaluated on a dataset with completely separated channel pairs' data.\footnote{This degree of degradation is expected and reasonable, as the model was tested on unseen data with different characteristics over channel links.}
These results are due to the attention mechanism focusing on target-relevant regions and delivering only the necessary information from the skip connections to the decoder for generating the segmentation map.
It is also important to note that this performance is achieved without transfer learning.

Next, we evaluate the tracking accuracy of the proposed method using three anchors. 
The results in Table VI show that the proposed method using the attention U-Net achieves the lowest errors across all metrics. 
Specifically, the proposed method reduces the 50th percentile error by 27.3\% compared to the CNN-based regression method and by 13.8\% compared to the MAP approach. 
Similarly, the 90th percentile error is reduced by 79.5 cm and 10.0 cm, respectively. 
As shown in Fig. 10(b), the proposed method demonstrates superior performance across the entire range. 
While the method using vanilla U-Net also outperforms the SOTA approaches, its performance for maximum errors is comparable to the MAP approach.

Fig. 10(a) visualizes the localization result of the proposed method, showing its ability to reasonably track the target's movements even under the constraints of limited anchors.

\section{Conclusion}
We have proposed a DFL system that uses COTS UWB transceivers to estimate the position of the target without the need for dedicated radar equipment. 
The system leverages a DL-assisted approach to enhance accuracy and reduce computational costs.

\begin{itemize}
    \item The variance profile of the CIR is analyzed using a 1-D U-Net DL model to predict the RoI, suppressing noise that interferes with localization.
    \item A hybrid system combining DL and a particle filter is developed for real-time localization, achieving an average processing time of 4~ms. 
    \item The proposed system shows high accuracy with an average RMSE of 15~cm and a 90th percentile error of 24~cm, demonstrating improved performance over existing SOTA methods.
\end{itemize}

The proposed system can be implemented without additional hardware installation using existing UWB infrastructure. 
For example, when operating simultaneously with the downlink time difference of arrival system, the CIRs from existing UWB anchors can be fully reused.

\section{Discussion}


To improve localization performance, monostatic radar measurements can be utilized. In practice, recent UWB transceivers not only provide ranging mode for device to device but also support monostatic radar mode. Since the Tx and Rx are synchronized in monostatic radar, the calculation of the CIR variance becomes straightforward, and Doppler frequency shift (DFS) can be extracted. This DFS corresponds to the radial velocity and can be used in combination with the existing CIR variance as part of the information fusion in the particle filter.

Future works will extend the system to handle multiple targets, improve robustness in cluttered environments, and achieve domain generalization for DL.

\end{document}